\documentclass[12pt]{article}
\usepackage{subeqn}
\usepackage{epsfig}
\newcommand{\be}{\begin{equation}}
\newcommand{\ee}{\end{equation}}

\newcommand{\no}{\noindent}
\newcommand{\ce}{\begin{center}}
\newcommand{\nc}{\end{center}}

\makeatletter
\@addtoreset{equation}{section}
\makeatother
\renewcommand{\theequation}{\thesection.\arabic{equation}}

\def\sqr#1#2{{\vcenter{\vbox{\hrule height.#2pt
\hbox{\vrule width.#2pt height#1pt \kern#1pt
\vrule width.#2pt} \hrule height.#2pt}}}}

\def\operp{\hbox{${\kern+.25em{\bigcirc}
\kern-.85em\bot\kern+.85em\kern-.25em}$}}

\def\lsim{\;\raise0.3ex\hbox{$<$\kern-0.75em\raise-1.1ex\hbox{$\sim$}}\;}
\def\gsim{\;\raise0.3ex\hbox{$>$\kern-0.75em\raise-1.1ex\hbox{$\sim$}}\;}
\def\no{\noindent}

\def\ce{\centerline}
\def\ve{\vfill\eject}
\def\rdots{\mathinner{\mkern1mu\raise1pt\vbox{\kern7pt\hbox{.}}\mkern2mu
\raise4pt\hbox{.}\mkern2mu\raise7pt\hbox{.}\mkern1mu}}

\def\e e{$e^+ e^-$ }

\begin{document}

\ce{\bf MASSES AND INTERACTIONS OF $q$-FERMIONIC KNOTS}

\vskip.5cm

\ce{\it Robert J. Finkelstein$^a$ and A. C. Cadavid$~^b$}
\vskip.3cm

\ce{$^a$Department of Physics and Astronomy}
\ce{University of California, Los Angeles, CA 90095-1547}
\vskip0.3cm
\ce{$^b$Department of Physics}
\ce{California State University, Northridge, CA  91330}
\vskip1.0cm

\no{\bf Abstract.} The $q$-electroweak theory suggests a description of
elementary particles as solitons labelled by the irreducible representations of $SU_q(2)$.
Since knots may also be labelled by the irreducible representations of
$SU_q(2)$, we study a model of elementary particles based on a
one-to-one correspondence between the four
families of Fermions (leptons, neutrinos, (-1/3) quarks, (2/3) quarks) and
the four simplest knots (trefoils).  In this model the three particles of each family are
identified with the ground and first two excited states of their common
trefoil.  Guided by the standard electroweak theory we calculate conditions
restricting the masses of the fermions and the interactions between them.

In its present form the model predicts a fourth generation of fermions as well
as a neutrino spectrum. The same model with $q\cong 1$ is compatible with the Kobayashi-Maskawa matrix. 
Depending on the test of these predictions, the model
may be refined.

\vskip2.0cm

\no {\bf UCLA/05/TEP/20} 

\ve

\section{Introduction.}

We continue to investigate the possibility of describing the elementary fermions as
knotted solitons.${^1}$  These knots may be understood either as simply
symbols (labels of particles) or as real physical structures such as
knotted flux tubes.  To relate the simplest particles to the simplest knots,
we represent each of the 4 families of elementary fermions by a separate
soliton labelled by one of the 4 possible trefoils, e.g. the family
$e,\mu,\tau$ is represented by a single trefoil, while the $e,\mu$, and
$\tau$ particles are separately identified as different states of excitation
of their common trefoil.  In this paper we attempt to calculate interactions
between these $q$-fermions mediated by the $q$-gauge vector, or alternatively to determine the $q$-currents.

\vskip.5cm

\section {The Origin of the Knots.}

Our work is based on the possibility that $SU_q(2)$ is an effective
phenomenological symmetry.  If it is, the symmetry group of the standard
electroweak theory may be regarded as a degenerate form of $SU_q(2)$.  The linearized form of the theory based on the $q$-symmetry
is indeed in approximate agreement with the standard theory in lowest
order.$^2$

To go beyond the linearization one may expand the quantum fields in
irreducible representations $(D^j_{mm^\prime}(q|a,\bar a,b,\bar b))$ of
$SU_q(2)$ where the arguments $(a,\bar a,b,\bar b)$ obey the algebra
of $SU_q(2)$.  Then the normal modes, besides describing states of 
momentum and spin, will also contain factors $D^j_{mm^\prime}(q|a,\bar a,
b,\bar b)$.  These are polynomials in the non-commuting arguments
$(a,\bar a,b,\bar b)$ with eigenstates $|n\rangle$.  Since the different
normal modes therefore have internal excited states, they may be described
as solitons ($q$-solitons) rather than as point particles.  A class of these normal modes may
be related to knots and labelled by $D^{N/2}_{\frac{w}{2}\frac{r+1}{2}}
(q|a,\bar a,b,\bar b)$ where $(N,w,r)$ mean the number of crossings, the
writhe, and the rotation of the knot.${^1}$  (To correctly represent a knot
the three integers $(N,w,r)$ must satisfy certain knot constraints, e.g.
$w$ and $r$ must be of opposite parity.)

\vskip.5cm

\section {Representation of the Elementary Particles.}

We now propose that the elementary particles may be usefully labelled by
the irreducible representations of $SU_q(2)$ in the form $D^{N/2}_{\frac{w}{2}
\frac{r+1}{2}}$.  Depending on whether $N$ is even or odd, we assume that
$D^{N/2}_{\frac{w}{2}\frac{r+1}{2}}$ represents either a boson or a
fermion respectively.

The lowest possible value of $N$ is 3 and the corresponding knot is a 
trefoil.  It is then natural to associate the elementary fermions with
the ground and lowest excited states of the trefoils.  There are 4 trefoils described
by
\be
(w,r) = (3,-2),~(3,2),~(-3,-2),~(-3,2)
\ee
There are also 4 families of elementary fermions, namely:
\be
(e,\mu,\tau),~(d,s,b),~(u,c,t),~(\nu_e,\nu_\mu,\nu_\tau)
\ee

The four trefoils and associated $D^{N/2}_{\frac{w}{2}\frac{r+1}{2}}$ are
shown in Fig. 1: 

\vskip1.0cm
\begin{figure}[h]
\begin{center}
\epsfig{file=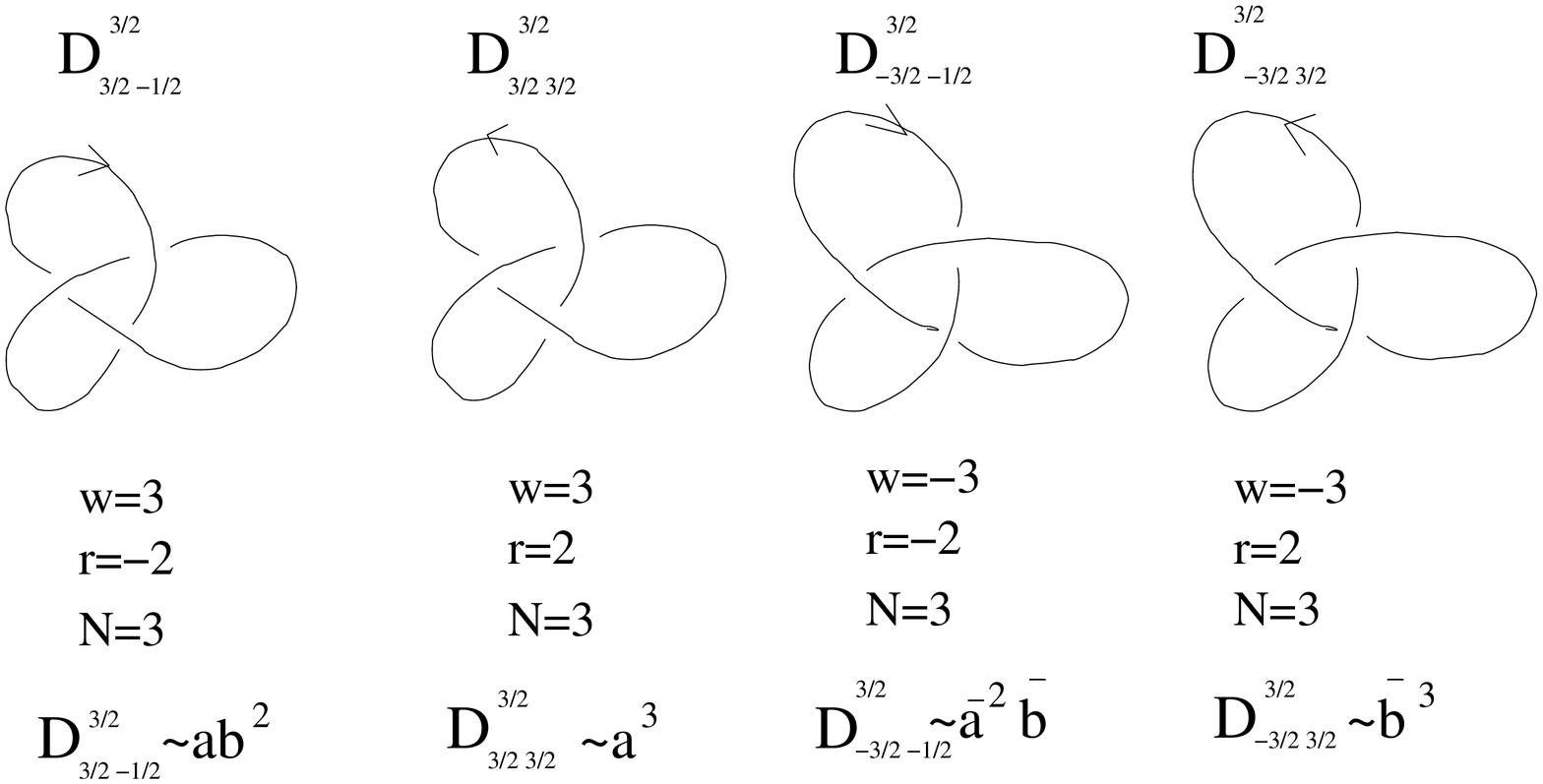,height=3.5in, width=6in}
\end{center}
\end{figure}
\vskip1.0cm
\begin{center}
{{\bf Figure 1.}}
\end{center}

For the last line of Fig. 1, see Eqs. (8.1)-(8.4).

Each of the 4 families of elementary fermions may be represented by one of
the 4 possible trefoils.  The 3 individual fermions belonging to a single
family are then assumed to represent 3 different states of excitation of a
single trefoil.

Members of the 4 families have the following values of $(t,t_3,Q)$, i.e. the
isotopic spin, its 3-component and the charge, and we shall 
tentatively assume that $w$
and $r$ (labelling the writhe and rotation of their common trefoil) have the
values shown in the same table:

\be
\begin{array}{ccccccc}
& \underline{t} & \underline{t_3} & \underline{Q} & \underline{w} & 
\underline{r} & \underline{D^{N/2}_{\frac{w}{2}\frac{r+1}{2}}}\\
(e,\mu,\tau) & 1/2 & -1/2 & -1 & 3 & 2 & D^{3/2}_{\frac{3}{2}\frac{3}{2}} \\
(\nu_e,\nu_\mu,\nu_\tau) & 1/2 & 1/2 & 0 & -3 & 2 & D^{3/2}_{-\frac{3}{2}
\frac{3}{2}} \\
(d,s,b) & 1/2 & -1/2 & -1/3 & 3 & -2 & D^{3/2}_{\frac{3}{2}-\frac{1}{2}} \\
(u,c,t) & 1/2 & 1/2 & 2/3 & -3 & -2 & D^{3/2}_{-\frac{3}{2}-\frac{1}{2}} \\
\end{array}
\ee 

\no The assignment of $w$ and $r$ to the 4 families is discussed in
paragraph 6 and in Ref. 1.

In the preceding table we have assumed the following relations between
conventional (point particle) labels and knot (soliton) labels for the
elementary fermions.

\be
\begin{array}{cccc}
\underline{e\mu\tau} & \underline{dsb} & \underline{uct} & \underline
{\nu_e\nu_\mu\nu_\tau} \\
t=\frac{N}{6} & t=\frac{N}{6} & t=\frac{N}{6} & t=\frac{N}{6} \\
t_3=-\frac{w}{6} & t_3=-\frac{w}{6} & t_3=-\frac{w}{6} & t_3=-\frac{w}{6} \\
Q=-\frac{1}{4}r-\frac{1}{2} & Q=\frac{1}{4}r+\frac{1}{6} &
Q=-\frac{1}{4}r+\frac{1}{6} & Q=-\frac{1}{4}r+\frac{1}{2} \\
\end{array}
\ee

\no These relations between $(t,t_3,Q)$ and $(N,w,r)$ define a knot model.
These linear relations satisfy (3.3).  This trial knot model
then establishes a unique match between the elementary fermion families and
the trefoils.

\section {Other Knots.}

Since the trefoils characterized by $(N=3,~w=\pm 3,~r=\pm2)$ are the
simplest knots, they have been chosen to represent the simplest particles:
the leptons and quarks.  One may obtain higher knots by forming a connected
sum of trefoils: These higher knots may be interpreted as bosonic or fermionic
depending on whether $N$ is even or odd.  In this way one may replicate the quark building up principle; then the mesons are two connected trefoils and the hadrons are three connected trefoils.

Any knot may be represented by $D^{N/2}_{\frac{w}{2}\frac{r+1}{2}}
(q|a,\bar a,b,\bar b)$ which is a $q$-polynomial, just as any algebraic curve may be
represented by a numerically valued polynomial.  On the other hand, not every
$D^j_{mn}(q|a,\bar a,b,\bar b)$ represents a knot; according to our ideas,
however, these non-knots still represent states of excitation of the field,
and their symbols, forming a complete orthogonal basis, would all be required in
the underlying field theory.

\section {Representation of $W^+W^-Z$ and $A$.}

Since the electroweak vector fields are responsible for pair production one
might try to think of the knots associated with these vectors as fusions
of the knots representing leptons or quarks.  Since we are associating these
elementary fermions with trefoils we shall represent the intermediate vectors 
as di-trefoils, as shown in the figures and tables.

\vskip1.0cm
\begin{figure}[h]
\begin{center}
\epsfig{file=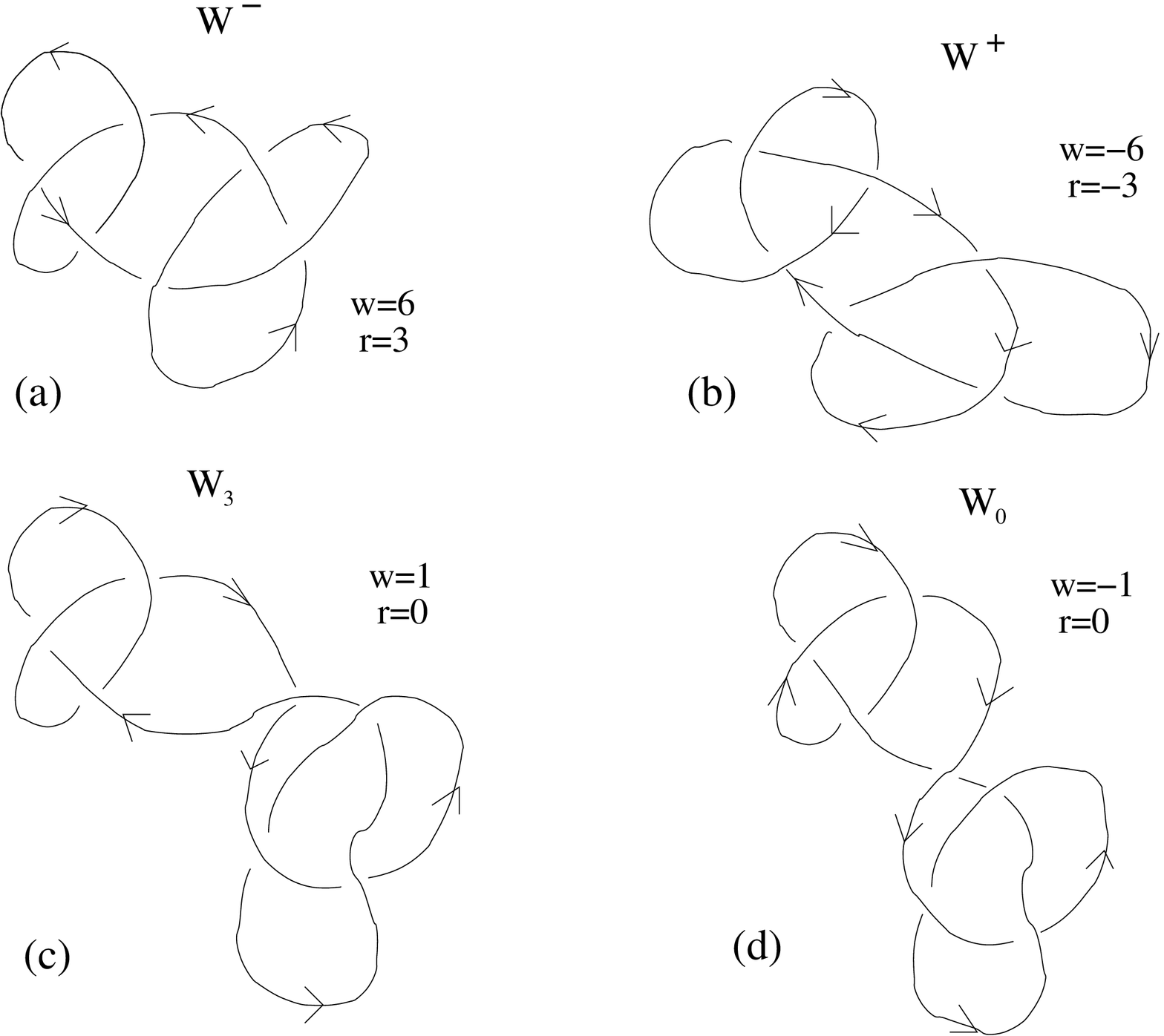,height=5in,width=6in}
\end{center}
\end{figure}
\vskip 4in
\noindent
{\bf Figure 2. Knot Representation of Gauge Vectors. The particle 
labelling is: (a) $ {\cal{D}}^{1}_{-1\;\;0}$ ; 
(b) ${\cal{D}}^{1}_{1\;\;0}$;
(c) ${\cal{D}}^{1}_{0\;\;0} $;
(d)$ {\cal{D}}^{1}_{-1\;\;1} $ }.

\vskip1.0cm

\begin{center}{
\begin{tabular}{l|ccccccc|c|c}
   & $t$ & $t_{3}$ &  $Q$ & $t_{0}$ & $N$ & $w$ & $r$ & ${\cal{D}}^{t}_{t_{3} \; t_{0}}$ \\
\hline
$ \begin{array}{c}
        W^{+} \\
        W^{-} \\
        W_{3} \\
        W_{0} \\            
     \end{array}$  & 
$\begin{array}{r}
           1\\
           1\\
           1\\
           1\\
    \end{array} $ &         
$\begin{array}{r}
           1\\
           -1\\
           0\\
           -1\\
    \end{array} $ &         
$\begin{array}{r}
           1\\
           -1\\
           0\\
           0\\
    \end{array} $ &         
$\begin{array}{r}
           0\\
           0\\
           0\\
           1\\
    \end{array} $ &         
$\begin{array}{r}
           6\\
           6\\
           7\\
           7\\
    \end{array} $ &         
$\begin{array}{r}
           -6\\
           +6\\
           1\\
           -1\\
    \end{array} $ &         
$\begin{array}{r}
           -3\\
           +3\\
           0\\
           0\\
    \end{array} $ &         
$\begin{array}{l}
          {\cal{D}}^{1}_{1\;\;0} \\
           {\cal{D}}^{1}_{-1\;\; 0}\\
          {\cal{D}}^{1}_{0\;\;0} \\
           {\cal{D}}^{1}_{-1\;\; 1}\\          
    \end{array} $              
\end{tabular}
}
\end{center}
\begin{center}

{{\bf Table 1.}}
\end{center}

\no In this scheme negative charge corresponds to counter-clockwise
rotation.  Then $r$ is positive for both $e^-$ and $W^-$.  Note
that ${\cal{D}}^t_{t_3t_0}$ exhibits the particle rather than the
knot labelling.

The linear relations between the quantum numbers $(t,t_3,Q)$ and the knot
integers $(N,w,r)$ are shown in Table 2.

\begin{center}{
\begin{tabular}{|c|c|c|}
$W^{+}$ and $W^{-}$  & $W_{3} $ & $W_{0}$ \\
\hline
$ \begin{array}{l}
            t=\frac{N}{6} \\
            t_{3}=-\frac{w}{6} \\
            Q=-\frac{1}{3}r \\
       \end{array} $ &
$  \begin{array}{l}
            t=\frac{N-1}{6} \\
            t_{3}=w-1 \\
            Q=r \\
       \end{array} $ &
$ \begin{array}{l}
            t=\frac{N-1}{6} \\
            t_{3}=w \\
            Q=r \\
       \end{array} $ 
\end{tabular}
}
\end{center}

\begin{center}
{{\bf Table 2.}}
\end{center}

Since the charge is proportional to the rotation of the knot in this scheme,
the component trefoils must have opposite rotations when the di-trefoil
represents a neutral vector.  If the two components do have opposite
rotations, however, there must be an additional crossing in the knot
diagram.  The knots representing $W_3$ and $W_0$ then differ in the writhe
of the crossing as shown.  The degeneracy between $W_3$ and $W_0$ is thus 
resolved by the differing values of the writhe, and in the standard theory
by coupling $W_0$ to $U(1)$ or by the introduction of the Weinberg angle.
Here we follow the standard theory by defining
\be
\begin{array}{rcl}
Z &=& -W_0\sin\theta + W_3\cos\theta \\
A &=& W_0\cos\theta + W_3\sin\theta \\
\end{array}
\ee
We have assumed that the number of intersections is (even, odd) for
(bosonic, fermionic) knots.  Although the number of intersections for
$W_0$ and $W_3$ separately is 7, this does not violate the (even, odd)
rule for the physical fields since $A$ and $Z$ are linear combinations of $W_0$ and $W_3$.  Hence the $A$ and $Z$ field quanta, being composite knots with 14 intersections, obey the (even, odd) rule.  In Table 2 we have arranged the relation between the isotopic spin and knot labels so
that all the di-trefoils lie in the same $SU_q(2)$ multiplet.

\section {Masses of Fermions.$^1$}

We follow the standard theory in assuming that the masses of the
fermions depend on the Higgs field $(\varphi)$ at the minima in the Higgs
potential.  The mass operator in the Hamiltonian density is then taken
to be
\be
{\cal{M}} = (\bar\psi_L\varphi\psi_R + \bar\psi_R\varphi\psi_L) 
\ee
\no Since $\psi_R$ is a singlet in the standard theory we assume that it
is also a singlet in the $SU_q(2)$ theory.  Then
\be
{\cal{M}} = \bar\psi_L\varphi + \varphi\psi_L
\ee
\no Now replace the fields $\psi_L$ and $\varphi$ by their normal
modes that represent trefoils.  We have been assuming that 
all fields including the Higgs field and
therefore the Higgs potential lie in the $q$-algebra. Let the
Higgs potential be chosen so that its minima
lie at the trefoil points.  The Higgs field at
these points is then
\be
\varphi = \rho(w,r) D^{3/2}_{\frac{w}{2}\frac{r+2}{2}}(a,\bar a~b,\bar b)
\ee
where $(w,r)$ is a trefoil point.  Then the mass operator (6.2) at 
$(w,r)$ becomes
\be
{\cal{M}}(w,r) = \rho(w,r)[\bar\psi_LD^{3/2}_{\frac{w}{2}\frac{r+1}{2}}
+ D^{3/2}_{\frac{w}{2}\frac{r+1}{2}}\psi_L]
\ee
\no and the mass operator associated with any soliton $(w^\prime,r^\prime)$
and Higgs $(w,r)$ becomes
\be
{\cal{M}}(w^\prime r^\prime;wr) = \rho(w,r)
\left[\bar D^{3/2}_{\frac{w^\prime}{2}\frac{r^\prime+1}{2}}D^{3/2}_{\frac{w}{2}\frac{r+1}{2}}
+ D^{3/2}_{\frac{w}{2}\frac{r+1}{2}} D^{3/2}_{\frac{w^\prime}{2}
\frac{r^\prime+1}{2}}\right]
\ee
\no Here we have dropped the multiplier of $D^{3/2}_{\frac{w}{2}
\frac{r+1}{2}}$ in the normal mode expansion that pertains to the momentum and spin of the fermions, since that factor would cancel
out in the following discussion.
The expectation value of ${\cal{M}}(w^\prime,r^\prime;w,r)$ vanishes unless
$w=w^\prime$ and $r=r^\prime$.  Then, since only the first term of
(6.5) contributes to the expectation value, we have
\be
\langle n|{\cal{M}}(w,r)|n\rangle =
\rho(w,r)\langle n|\bar D^{3/2}_{\frac{w}{2}\frac{r+1}{2}}
D^{3/2}_{\frac{w}{2}\frac{r+1}{2}}|n\rangle
\ee

To accommodate the 4 families one needs 4 minima in the Higgs potential.
These minima may be labelled by the magnitudes of the Higgs field $\varphi$
and by the associated Higgs trefoils.  The mass scale of each family is
determined by $\varphi$ at the minimum for that family, and the trefoil for
that family must agree with the trefoil for $\varphi$.  With this
understanding, Eq. (6.6) implies
\be
m_n(w,r) = \rho(w,r)\langle n|\bar D^{3/2}_{\frac{w}{2}\frac{r+1}{2}}
D^{3/2}_{\frac{w}{2}\frac{r+1}{2}}|n\rangle
\ee
\no where $m_n(w,r)$ is the mass of the $(w,r)$ soliton at the n$^{\rm th}$
level.  The different mass spectra corresponding to the different solitons
are by Ref. (1) or by (6.7), and (8.1) and (8.3) as follows:
\begin{subequations}
\begin{eqnarray}
\mbox{~~I} & & m_n(3,2) = \rho(3,2)
\Delta(\frac{3}{2},\frac{3}{2})(1-q^{2n-2}|\beta|^2)
(1-q^{2n-4}|\beta|^2)(1-q^{2n-6}|\beta|^2) \\
\mbox{~II} & & m_n(3,-2) = \rho(3,-2)\Delta(\frac{3}{2},-\frac{1}{2})
[q^{4n}|\beta|^4-q^{6n-2}|\beta|^6] \\
\mbox{III} & & m_n(-3,-2) = \rho(-3,-2)\Delta(-\frac{3}{2},-\frac{1}{2})
[q^{2n}|\beta|^2(1-q^{2n}|\beta|^2)
(1-q^{2n+2}|\beta|^2)] \\
\mbox{~IV} & & m_n(-3,2) = \rho(-3,2) \Delta(-\frac{3}{2},\frac{3}{2})
q^{6n}|\beta|^6 
\end{eqnarray}
\end{subequations}
\no Since all masses $m_n(w,r)$, within a single spectrum are proportional to
$\rho(w,r)\Delta(w,r)$, one may compute ratios of these masses without ambiguity.
In calculating these ratios we assume that only the three lowest states of
each soliton are occupied.  Set
\be
M = \frac{\langle 1|{\cal{M}}|1\rangle}{\langle 0|{\cal{M}}|0\rangle}
\qquad \mbox{and} \qquad 
m = \frac{\langle 2|{\cal{M}}|2\rangle}{\langle 1|{\cal{M}}|1\rangle}
\ee

There is an equation for both $M$ and $m$ in each spectrum I-IV.
These two equations may be rewritten for $q$ and $|\beta|^2$ as follows:
\begin{subequations}
\begin{eqnarray}
\mbox{~~I} \qquad \quad   \frac{m-1}{m-q^6}&=&q^2\frac{M-1}{M-q^6} \qquad |\beta|^2 = q^6\frac{M-1}{M-q^6} \\
\mbox{~II}  \qquad\quad  \frac{m-q^4}{m-q^6} &=& q^2\frac{M-q^4}{M-q^6} \qquad
|\beta|^2 = \frac{m-q^4}{m-q^6} \\
\mbox{III}  \qquad\quad \frac{m-q^2}{m-q^6} &=& q^2\frac{M-q^2}{M-q^6} \qquad
|\beta|^2 = \frac{M-q^2}{M-q^6} \\
\mbox{~IV}  \qquad\quad \qquad M &=& m=q^6
\end{eqnarray}
\end{subequations}

\no The empirical input depends on the masses of the elementary fermions.
These are well determined for the leptons $(e,\mu,\tau)$, but for the
quarks they are not even well defined.  Since the quarks do not exist as
free particles, the quoted masses depend on the theoretical procedure for
defining them.  There is then a range of ``masses" given by the Particle
Data Group.$^3$

To solve the above equations for $q$ and $|\beta|^2$, we have chosen the
following values for $M$ and $m$.
\be
\begin{array}{lcc}
 &  \underline{M} & \underline{m} \\
(1) \qquad \qquad e\mu\tau & 193 & 16.7\\
(2) \qquad \qquad dsb & 37.5 & 31.8 \\
(3) \qquad \qquad uct & 750 & 117 \\
(4) \qquad \qquad \nu_{e}, \nu_{\mu}, \nu_{\tau} & ? & ? \\
\end{array}
\ee

These ratios are based on the masses of the quarks recorded here:
\be
\begin{array}{ccccccc}
u & d & c & s & t & b & \\
.002 & .004 & 1.5 & .15 & 176 & 4.77 & \mbox{GeV/c}^2
\end{array}
\ee

One may try to match the four familiies (1)-(4) shown in (6.11) with the
four spectra (I-IV) shown in (6.8).  It is clear that none of the
three families (1), (2), (3) match (IV).  Therefore we assign (IV) to the
neutrino family.  Next write (6.10a)-(6.10c) as algebraic equations
in $q^2$ and assign the equation of lowest degree to the lepton family
(since the leptons do not have hypercharge or gluon charge.$^1$)
Then if we assign the I, II, and III spectra to $(e,\mu,\tau),
(d,s,b)$ and $(u,c,t)$ respectively, we find that the roots of (6.10a)-
(6.10c) where $q$ is closest to unity are
\be
(e\mu\tau)  \quad  q = 1.46 \quad |\beta| = 3.20 
\ee
\be
(dsb) \quad q = 1.76 \quad |\beta| = 3.35 
\ee
\be
(uct) \quad q = 2.14 \quad |\beta| = 1.07 
\ee

It also turns out that any other match is also good, i.e., if
$q=q(M,m,w,r)$, then it is found that $q$ depends mainly on $M$ and $m$, and
is nearly independent of $w$ and $r$.  We shall not, however, represent each
of the three fermion families as a linear combination of the three
trefoils, since they are topologically distinct, and consequently
there is a topological obstruction to any dynamical transition
between any two of them.  Therefore in this simplified model
we associate each family with a
single trefoil, or equivalently with a single normal mode or irreducible
representation, $D^{3/2}_{\frac{w}{2},\frac{r+1}{2}}$, where $(w,r)$
characterizes the trefoil. To match the families with the trefoils in a
unique way, we tentatively postulate the knot model described by (3.4).

Higher knots designated by $D^{N/2}_{\frac{w}{2}\frac{r+1}{2}}$ with the
same $(w,r)$ and $N>3$ are topologically equivalent and can therefore
dynamically decay to trefoils.  Moreover, if the dynamics requires that
lower $N$, as well as lower $n$, implies lower energy, then only the 
trefoil solitons will be stable and recognizable as particles.  (The
topologically equivalent but dynamically unstable higher knots differ
from the trefoils by a connected sum of curls.)

Eqs. (6.8a) through (6.8d) are of the form
\be
m_n(w,r) = \rho(w,r) F(w,r;n,q,\beta)
\ee
\no By (6.13), (6.14), (6.15) one sees that $F(w,r;n,q,\beta)$ is negative
in (6.8a) and  (6.8b), but it is positive in (6.8c) and (6.8d).  Therefore the
first two minima $(\rho(3,2), \rho(3,-2))$, must be
negative while $\rho(-3,-2)$ and
$\rho(-3,2)$ must be positive to ensure that all masses
$m_n(w,r)$ are positive.

The magnitude of $\rho(w,r)$ sets the energy scale and differs for each
family.  The choice of $\rho(w,r)$ and $F(w,r;n,q,\beta)$ for each
family is determined by the knot model, i.e. by the postulated linear
relation between $(t_3,Q)$ and $(w,r)$ in (3.3) or (3.4), as well as by
the postulated relation between knots and the irreducible representations
of $SU_q(2)$, namely $D^{N/2}_{\frac{w}{2}\frac{r+1}{2}}$.

The value of $q$ depends only weakly on $(w,r)$ in $F(w,r;n,q,\beta)$ but
it does depend strongly on $\beta$ and $m_n(w,r)$.  The parameter $q$ therefore
behaves like a running coupling constant, where $\beta$ and 
$m_n(w,r)$ fix the energy scale.

We may interpret the numerical value of $q$ as a measure of the influence
of the fields that play a role in the determination of the fermionic masses
and that are excluded from the standard electroweak theory.  Consistent with this
view, $q$ is not far from unity; and the lepton family, having no gluon
charge, has a $q$ value closer to unity than the quark families.

We have assumed that the three observed particles of each family occupy the 3
lowest states of the soliton representing that family.  The model also permits
higher excited states but if these lie at very high energies, they may have such
short lifetimes that they would not be observable as particles.  The tentative
assignment that we have assumed in (3.3) leads to a fourth generation of
(-1/3 quarks) at $30m_b \sim 144$ GeV and a fourth generation of 
(2/3 quarks) at $100m_t \sim 17,600$ GeV.  The corresponding fourth generation
lepton would appear at $12m_\tau \cong 21.3$ GeV but is excluded
by the known decays of the $Z^0$.$^7$  If the assignments of $dsb$ and $uct$ are interchanged so that $dsb$ corresponds to III and $uct$ to II then
the fourth generation would appear at $30.4m_b$ and $102m_t$.  If a fourth
generation should be observed then a unique assignment of the $(dsb)$ and $(uct)$ families to trefoils could be put on an empirical basis.  In any case further
refinements of the model would depend on whether any or none of the fourth
generation particles is observed.  The predicted neutrino spectrum is a
further test of the model.  The neutrino data are very sparse, but are compatible
with $q\cong 1$, leading by (6.8d) to a geometric hierarchy of nearly equal
masses.$^4$

\section {Interactions Mediated by a Vector Field.}

We are next interested in interactions that stem from the gauge invariant
terms
\be
\bar\psi\nabla\!\!\!\!/\psi
\ee
\no where $\nabla\!\!\!\!/ = \gamma^\mu\nabla_\mu$ is the gauge covariant
derivative.  This term gives rise to
\be
\bar\psi W\!\!\!\!/\psi
\ee
\no where
\be
W = \nabla-\partial
\ee

When $\psi$ and $W$ are expanded in absorption and emission operators, the
normal modes will specify momentum, spin and species of soliton; in more
detail it will specify the internal state of the soliton.  To describe the
interaction between fermions mediated by a vector particle one replaces
the field operators by normal modes.  Schematically
\be
\bar\psi W\!\!\!\!/\psi \to \bar D^\alpha_i W\!\!\!\!/ D^\beta_j
\ee
\no where
\be
D^\alpha_i = D^\alpha|i\rangle
\ee
\no and $|i\rangle$ is an ``internal" state, like a spin state.  Here
$\alpha$ runs over the 4 kinds of trefoils, i.e., $\alpha$ fixes
$(w,r)$ while $|i\rangle$ labels the particle and the level of the
trefoil spectrum.  Hence
\be
\bar\psi W\!\!\!\!/\psi \to\langle i|\bar D^{3/2}_{\frac{w_1}{2}\frac{r_1+1}{2}}
W\!\!\!\!/ D^{3/2}_{\frac{w_2}{2}\frac{r_2+1}{2}}|j\rangle
\ee
\no where we have abstracted just the part of the matrix element that depends
on the $q$-algebra.

In (7.6) $W$ is to be replaced by a normal mode or by a linear combination
of normal modes.  Lacking a firm {\it a priori} basis, this choice must
b‮e determined by empirical data.  The problem here is similar to that faced
in the earlier days of weak interaction theory where various linear
combinations of the five fundamental forms were proposed before the 
decisive experiment requiring V-A was performed.  Here we shall be guided
on the one hand by the di-trefoil construction (Fig. 2) and on the other by
the experimental requirement that each lepton be pair produced with only
its ``own" neutrino (lepton conservation) as well as by the additional
restriction usually expressed as the universal Fermi interaction.

To satisfy these requirements we have made the following choices:
\be
\begin{array}{lcccc}
& \; \underline{W^-} & \; \underline{W^+} & \; \underline{W^3} & \; \underline{W^0} \\
(a) & \; D^3_{30} & \; D^3_{-30} & \; D^3_{00} & \; D^3_{-11} \\
(b) & \; D^{N/2}_{+\frac{w}{2}\frac{r-3}{2}} & \; 
D^{N/2}_{+\frac{w}{2}\frac{r+3}{2}} & \; 
D^{(N-1)/2}_{w-1,r} & \; 
D^{(N-1)/2}_{w,r+1} \\
\end{array}
\ee

Line (a) is chosen so that there is no change in level between initial and
final states and therefore the usual fermion pairs are produced by 
$W$.

Line (b) is a relabelling of line (a) in terms of the knot signature
$(N,w,r)$ according to Tables 1 and 2.  A more general possibility is
\be
C_-(q,\beta)D^3_{30}W^-+C_+(q,\beta)D^3_{-30}W^++C_3(q,\beta)
D^3_{00}W^3+C_0(q,\beta)D^3_{-11}W^0
\ee
\no Since we are interested mainly in relative rates in this paper, we
shall usually not be concerned with the possible $(q,\beta)$ dependence of the four coefficients.

The representatives of the charged vectors, namely $D^3_{30}$ and
$D^3_{-30}$, are conjugate monomials (up to $-q_1^3$) while the
corresponding representatives of the neutral vectors, namely
$D^3_{00}$ and $D^3_{-11}$, are polynomials in the $(b,\bar b)$
subalgebra.  These polynomials must satisfy the requirement that
$W^3$ and $W^0$ have non-vanishing matrix elements between
neutrino states.  It follows that $D^3_{00}$ and $D^3_{-11}$ must
lie in the $(b,\bar b)$ subalgebra since the neutrino states
lie in this subalgebra.  By (8.1) the conditions that $D^j_{mm^\prime}$
lies in the $(b,\bar b)$ algebra are
\be
s+t = j-m
\ee
\no and
\be
s+t = j+m^\prime
\ee
\no Hence
\be
m+m^\prime = 0
\ee

By (8.1) the condition that $D^j_{mm^\prime}$ be a function of only
the product $(b\bar b)$ requires in addition to (7.9) and (7.10)
\be
s+t = j+m
\ee
\no By (7.10) and (7.12)
\be
m=m^\prime
\ee
\no Both $D^3_{00}$ and $D^3_{-11}$ satisfy (7.11).  Only $D^3_{00}$
satisfies (7.13) as well.

The neutral sector of the algebra is then determined by the
neutral vectors and the neutrinos to be the $(b,\bar b)$ subalgebra.

\section {The ``Internal" Modes.}

To evaluate (7.6) one expresses the irreducible representations of
$SU_q(2)$ as follows:$^1$
\be
\begin{array}{rcl}
D^j_{mm^\prime}(a,\bar a,b,\bar b) &=& \Delta^j_{mm^\prime} \sum_{s,t}
\left\langle\matrix{n_+ \cr s \cr}\right\rangle_1
\left\langle\matrix{n_- \cr t}\right\rangle_1
q_1^{t(n_++1-s)}(-1)^t\delta(s+t,n^\prime_+) \\
& &\times a^sb^{n_+-s}\bar b^t\bar a^{n_--t} \\
\end{array}
\ee
\no where
\be
\begin{array}{rcl}
n_\pm &=& j\pm m \\ n^\prime_\pm &=& j\pm m^\prime \\
\end{array} \quad
\left\langle\matrix{n \cr s \cr}\right\rangle_1 =
{\langle n\rangle_1!\over \langle s\rangle_1!\langle n-s\rangle_1!}
\quad \langle n\rangle_1 = {q_1^{2n}-1\over q_1^2-1} \nonumber
\ee 
\be
\Delta^j_{mm^\prime} = \left[{\langle n^\prime_+\rangle_1!~
\langle n^\prime_-\rangle_1!\over
\langle n_+\rangle_1!~\langle n_-\rangle_1!}\right]^{1/2} \qquad
q_1 = q^{-1} \nonumber
\ee

The special cases (3.3) and (7.7a) when written out according to (8.1) are
\vskip.3cm

\no {\underline{\bf Fermions}}
\begin{subequations}
\be
\begin{array}{ccccc}
(w,r) & (3,2) & ~~~(3,-2)~~~ & ~~~(-3,-2)~~~ & ~~~(-3,2) \\
& D^{3/2}_{\frac{3}{2}\frac{3}{2}} & D^{3/2}_{\frac{3}{2}-\frac{1}{2}} &
D^{3/2}_{-\frac{3}{2}-\frac{1}{2}} & D^{3/2}_{-\frac{3}{2}\frac{3}{2}} \\
& & & & \\
& a^3 & \Delta^{3/2}_{\frac{3}{2}-\frac{1}{2}}\left\langle\matrix{3\cr
1\cr}\right\rangle_1ab^2 & -\Delta^{3/2}_{-\frac{3}{2}-\frac{1}{2}}
\left\langle\matrix{3\cr 1\cr}\right\rangle_1q_1\bar b\bar a^2 & 
-q_1^3\bar b^3 \\
& & & & \\
& (e\mu\tau) & (dsb) & (uct) & (\nu_e\nu_\mu\nu_\tau) \\
\end{array}
\ee
\no where
\be
\Delta^{3/2}_{-\frac{3}{2}\frac{1}{2}}\left\langle\matrix{3 \cr 1\cr}
\right\rangle_1 = \langle 3\rangle_1^{1/2}
\ee
\end{subequations}
\no To pass from particle to anti-particle we propose to take not only
the usual charge conjugation operator, but in addition to take the
$q$-conjugate as well.  The $q$-antifermions are represented by the adjoint
symbols, e.g., the $(\bar e\bar\mu\bar\tau)$ family is represented by
$\bar a^3$.
\vskip.3cm

\no {\underline{\bf Vectors}}
\begin{subequations}
\be
\begin{array}{cccc}
\underline{W^-} & \underline{W^+} & \underline{W^3} & \underline{W^0} \\
D^3_{30} & D^3_{-30} & D^3_{00} & D^3_{-11} \\
\left\langle\matrix{6 \cr 3 \cr}\right\rangle_1^{1/2} a^3b^3 &
-\left\langle\matrix{6 \cr 3\cr}\right\rangle_1^{1/2} q_1^3\bar b^3\bar a^3 &
f_3(b\bar b) & f_0(b,\bar b)  \\
\end{array}
\ee
\no where
\be
\Delta_{30}^3\left\langle\matrix{6\cr 3\cr}\right\rangle_1 =
\left\langle\matrix{6\cr 3\cr}\right\rangle_1^{1/2}  
\ee
\end{subequations}
\no and
\be
f_0(b,\bar b) = \left[\left\langle\matrix{4 \cr 2 \cr}\right\rangle_1
q^2(1-b\bar b)(1-q^2b\bar b)-\langle 2\rangle_1\langle 4\rangle_1 q^2_1(1-b\bar b)
(b\bar b) + q_1^{12}(b\bar b)^2\right]\bar b^2
\ee
\be
f_3(b\bar b) = \prod^2_{s=0}(1-q^{2s}b\bar b)-q^2\langle 3\rangle_1^2(b\bar b) \prod^1_{s=0}(1-q^{2s}b\bar b) + q_1^2\langle 3\rangle_1^2
(b\bar b)^2(1-b\bar b)-q_1^{12}(b\bar b)^3
\ee
\no Note that flavor changing neutral currents are absolutely
forbidden.  Note also that
\[
\bar D^3_{00} = D^3_{00}
\]
\[
\bar D^3_{-11}(b,\bar b) = D^3_{-11}(\bar b,b)
\]
\no In reducing (7.6) the following relations are useful:$^{1,5}$
\be
\begin{array}{ll}
ab = qba &  \qquad a\bar a+b\bar b = 1  \\
a\bar b = q\bar ba & \qquad \bar{a}a + q_1^2\bar{b}b = 1 \\
b\bar b = \bar bb & \\
\end{array}
\ee
\be
\begin{array}{rcl}
\bar b|n\rangle &=& q^n\beta^\star|n\rangle \\
b\bar b|n\rangle &=& q^{2n}|\beta|^2|n\rangle \\
\end{array}
\ee
\be
\begin{array}{rcl}
a|n\rangle &=& \lambda_n|n-1\rangle \\
|\lambda_n| &=& (1-q^{2(n-1)}|\beta|^2)^{1/2} \\
\end{array}
\ee
\be
\begin{array}{rcl}
\bar a|n\rangle &=& \mu_n|n+1\rangle \\
|\mu_n| &=& (1-q^{2n}|\beta|^2)^{1/2} \\
\end{array}
\ee
\be
\langle n|m\rangle = \delta(n,m)
\ee
\be
\begin{array}{rcl}
\bar a^n a^n &=&\displaystyle{\prod^{n-1}_{s=0}} (1-q_1^{2(s+1)}b\bar b) =
\displaystyle{\prod^n_{t=1}}(1-q_1^{2t}b\bar b) \\
a^n\bar a^n &=& \displaystyle{\prod^{n-1}_{s=0}}(1-q^{2s}b\bar b) \\
\end{array}
\ee

In the following we shall determine the dependence of the matrix
elements on the $q$-algebra.

\section {Lepton-Neutrino Couplings.}

\no (a) {\it Mediated by $W^-$:}

\be
\bar\ell(j) + W^- \to \bar\nu(i)\qquad \mbox{or} \qquad
W^- \to\ell(j) + \bar\nu(i)
\ee
\no The matrix element for the absorption of a $\bar\ell(j)$ and the
emission of a $\bar\nu(i)$ is by (8.4) and (8.5)
\be
\begin{array}{rcl}
m(i,j) &=& \langle i|\stackrel{=}{D}^{3/2}_{-\frac{3}{2}\frac{3}{2}}
D^3_{30}\bar D^{3/2}_{\frac{3}{2}\frac{3}{2}}|j\rangle \\
&=&\left\langle\matrix{6\cr 3\cr}\right\rangle_1^{1/2}
q_1^3\langle i|(-\bar b^3)(a^3b^3)(\bar a^3)|j\rangle \\
\end{array}
\ee
\no where the double bar signifies an antiparticle in the final state
and where by (8.8) and (8.13)
\be
\begin{array}{rcl}
\bar b^3a^3b^3\bar a^3 &=& q^9(\bar bb)^3 a^3\bar a^3 \\
&=& q^9(\bar bb)^3(1-b\bar b)(1-q^2b\bar b)(1-q^4b\bar b) \\
\end{array}
\ee
\no Then
\be
m(i,j) = -\left\langle\matrix{6 \cr 3\cr}\right\rangle^{1/2}_1
q^{6+6n_i}|\beta|^6 f(n_i)f(n_i+1)f(n_i+2)\delta(i,j)
\ee
\no where
\be
f(n) = 1-q^{2n}|\beta|^2
\ee

The ratio of matrix elements at level $(n+1)$ to those at level $(n)$ is
\be
\begin{array}{rcl}
R_n =\frac{m(n+1)}{m(n)} &=& q^6 \frac{f(n+1)f(n+2)f(n+3)}
{f(n)f(n+1)f(n+2)} \\
&=& q^6 \frac{1-q^{2n+6}|\beta|^2}{1-q^{2n}|\beta|^2} \\
\end{array}
\ee
\no Then
\be
\begin{array}{rcl}
M &\equiv& R_0 = q^6 \frac{1-q^6|\beta|^2}{1-|\beta|^2} \\
m &\equiv& R_1 = q^6 \frac{1-q^8|\beta|^2}{1-q^2|\beta|^2} \\
\end{array}
\ee

The Eqs. (9.7) may be rewritten as two equations for
$|\beta|^2$, namely:
\be
\begin{array}{rcl}
|\beta|^2 &=& \frac{M-q^6}{M-q^{12}} \\
|\beta|^2 &=& q^{-2} \frac{m-q^6}{m-q^{12}} \\
\end{array}
\ee
\no By eliminating $|\beta|^2$ one finds
\be
x^9 + m\langle 2\rangle_x x^4-M\langle 4\rangle_x x^3 + Mm = 0
\ee
\no where $\langle 2\rangle_x$ and $\langle 4\rangle_x$ are basic
numbers $\left(\langle n\rangle_x = \frac{x^n-1}{x-1}\right)$ and
\be
x = q^2
\ee

If we assume that the universal Fermi interaction that holds for point
particles in the standard theory also holds here, then
\be
M = m = 1
\ee
\no and (9.8) and (9.9) imply
\be
q=1
\ee
\no This result differs sharply from the results of (6.13)-(6.15)
and is a consequence of postulating lepton conservation as well as the
universal Fermi interaction $(M=m=1)$ that holds for point particles.
Depending on the extent that the U.F.I. may be violated between solitons,
one would find solutions of (9.9) differing from but close to unity.  

If $q$ is exactly unity, then $|\beta|=\frac{1}{2}\sqrt{2}$ by (9.8).  If $m$ and $M$ differ from unity, there will be corresponding
shifts in $(q,\beta)$ according to (9.8) and (9.9).

\vskip.3cm

\no (b) {\it Mediated by $W^+$:}

Now
\be
\ell(j) + W^+ \to \nu(i) \qquad \mbox{or} \qquad
W^+ \to \nu(i) + \bar\ell(j)
\ee

The matrix element for this reaction is
\be
\begin{array}{rcl}
m(i,j)^\prime &=& \langle i|\bar D^{3/2}_{-\frac{3}{2}\frac{3}{2}}
D^3_{-30}D^{3/2}_{\frac{3}{2}\frac{3}{2}}|j\rangle \\
&=& \langle i|(-q_1^3 b^3)\left(-\left\langle\matrix{6 \cr 3\cr}\right\rangle^{1/2}_1
q_1^3\bar b^3\bar a^3\right) a^3|j\rangle \\
&=& \left\langle\matrix{6\cr 3\cr}\right\rangle_1^{1/2} q^{6n_i-6}
|\beta|^6 f(n_i-1)f(n_i-2)f(n_1-3)\delta(i,j) \\
\end{array}
\ee
\no where $f(n)$ is given by (9.5).  Then
\be
R = \frac{m(i,j)^\prime}{m(i,j)} = -q^{-12}
\frac{f(n-1)f(n-2)f(n-3)}{f(n)f(n+1)f(n+2)}
\ee
\no is the ratio of the matrix elements for the two charge conjugate
reactions (9.1) and (9.13), up to the factor $C_+(q,\beta)/C_-(q,\beta)$.

Since $R$ is empirically very close to unity, (9.15) suggests that $q$ is
again very close to unity.  Hence the charge conjugate symmetry as well as
the universality of the Fermi interaction both imply that $q$ is near unity
in the interaction of leptons and neutrinos.  Therefore 
we conclude that the additional
degrees of freedom associated with masses of the leptons and neutrinos are
not excited in their pair production.  These last remarks depend on the
choice of $C_+(q,\beta)$ and $C_-(q,\beta)$ that in turn
are restricted by the relative masses of the vectors to be discussed
later.

We may take the view that the internal $SU_q(2)$ algebra is an effective
deformation of $SU(2)$ that depends on the background: in the case of the
soliton spectra the deviations of $q$ from unity are relatively large
but in the case of lepton-neutrino interactions, these deviations are
suppressed, just as they would be if we were dealing with point particles
rather than solitons, i.e. as if a weak charge were concentrated at the
center of an approximately spherically symmetric soliton.

\section {Charge Changing Quark Couplings.}

We first consider
\be
Q\left(j,-\frac{1}{3}\right) + W^+ \to Q\left(i,\frac{2}{3}\right)
\ee
\no where $Q\left(j,-\frac{1}{3}\right)$ is any quark of charge -1/3 and
$Q\left(i,\frac{2}{3}\right)$ is any quark of charge 2/3.

The matrix element for this process is by (8.4) and (8.5)
\be
\begin{array}{rcl}
m\left[\left(-\frac{1}{3} j\right) \to \left(\frac{2}{3} i\right)\right]
&=& \langle i|\bar D^{3/2}_{-\frac{3}{2}-\frac{1}{2}}D^3_{-30}
D^{3/2}_{\frac{3}{2}-\frac{1}{2}}|j\rangle \\
&=& {\cal C} \Delta^3_{30}\left\langle\matrix{6\cr 3\cr}\right\rangle_1 q_1^4
\langle i|(a^2b)(\bar b^3\bar a^3)(ab^2)|j\rangle \\
\end{array}
\ee
\no with 
\[ {\cal C} = \Delta^{3/2}_{\frac{3}{2}-\frac{1}{2}} 
\Delta^{3/2}_{-\frac{3}{2}-\frac{1}{2}} 
\left( \left\langle\matrix{3\cr 1\cr}\right\rangle_1\right)^2 \]

\no and where
\be
\begin{array}{rcl}
\langle i|(a^2b)(\bar b^3\bar a^3)(ab^2)|j\rangle &=&
q^8\langle i|(b\bar b^3)(a^2\bar a^3a)b^2|j\rangle \\
&=& q^8\langle i|(b\bar b^3)(a^2\bar a^2)(\bar aa)b^2|j\rangle \\
\end{array}
\ee
\no Then
\be
m\left[\left(-\frac{1}{3} j\right) \to \left(\frac{2}{3} i\right)\right]
= q^4 {\cal C} \left\langle\matrix{6\cr 3\cr}\right\rangle_1^{1/2} q^{6n_i}
|\beta|^6 f(n_i)f(n_i+1)f(n_i-1)\delta(i,j) 
\ee
\no where $f(n)$ is defined by (9.5).

The ratio of matrix elements at level $n+1$ to those at level $n$ is by
(10.4)
\be
R_n = q^6 \frac{1-q^{2(n+2)}|\beta|^2}{1-q^{2(n-1)}|\beta|^2} 
\ee
\no In particular
\be
\begin{array}{rcl}
R_0 &=& q^6 \frac{1-q^4|\beta|^2}{1-q^{-2}|\beta|^2} =
\frac{m(s+W^+\to c)}{m(d+W^+\to u)} \\
R_1 &=& q^6 \frac{1-q^6|\beta|^2}{1-|\beta|^2} =
\frac{m(b+W^+\to t)}{m(s+W^+\to s)} \\
\end{array}
\ee
\no Set
\be
\begin{array}{rcl}
M &=& R_0 \\
m &=& R_1 \\
\end{array}
\ee
\no Then by (10.6)
\be
\begin{array}{rcl}
|\beta|^2 &=& q^2\frac{M-q^6}{M-q^{12}} \\
|\beta|^2 &=& \frac{m-q^6}{m-q^{12}} \\
\end{array}
\ee
\no By the preceding equations for $|\beta|^2$
\be
q^2\frac{M-q^6}{M-q^{12}} = \frac{m-q^6}{m-q^{12}}
\ee
\no or
\be
q^{18} + mq^8\langle 2\rangle_{q^2}- Mq^6\langle 4\rangle_{q^2}
+ Mm = 0
\ee
\no Again if the matrix elements are equal for the following processes:
\be
\begin{array}{rcl}
d+W^+ &\to& u \\ s+W^+ &\to& c \\ b+W^+ &\to& t \\
\end{array}
\ee
\no we may set
\begin{subequations}
\be
M = m = 1
\ee
\no then (10.8) and (10.10) imply
\be
q=1 \qquad |\beta|= \frac{1}{2} \sqrt{2} = .707
\ee
\end{subequations}

Since the diagonal elements of the Kobayashi-Maskawa matrix are not quite equal however
(i.e. not strictly independent of $n$), Eq. (10.12a) is not exactly satisfied so that
$q$ and $|\beta|$ must differ slightly from (10.12b).

Let us next consider processes mediated by $W^-$:
\be
Q\left(\frac{2}{3},j\right) + W^-\to Q\left(-\frac{1}{3},i\right)
\ee
\no For this reaction (7.6) becomes
\be
- {\cal C} \left\langle\matrix{6\cr 3\cr}\right\rangle_1^{1/2}
q_{1} \langle i|\overline{ab^2}\cdot
a^3b^3\cdot\bar b\bar a^2|j\rangle = 
- {\cal C} \left\langle\matrix{6\cr 3\cr}\right\rangle_1^{1/2}
q_{1} \langle i|\bar b^2(\bar aa^3)(b^3\bar b)\bar a^2|j\rangle
\ee
\no where
\be
\begin{array}{rcl}
\overline{b^2}(\bar aa^3)(b^3\bar b)\bar a^2 &=& q^6(\bar b^2b^3)(\bar aa^3)\bar b\bar a^2 \\
&=& q^8\bar b^3b^3\bar aa^3\bar a^2 \\
&=& q^8(b\bar b)^3(\bar aa)(a^2\bar a^2) \\
&=& q^8(b\bar b)^3(1-q_1^2\bar bb)(1-b\bar b)(1-q^2b\bar b) \\
\end{array}
\ee
\no Then the matrix element (10.14) is by (10.15)
\be
\begin{array}{rcl}
m\left[\left(\frac{2}{3}~j\right)\to\left(-\frac{1}{3}~i\right)\right] &=&
- {\cal C} \left\langle\matrix{6\cr 3\cr}\right\rangle_1^{1/2} q^7
\langle i|(b\bar b)^3(1-q_1^2\bar bb)(1-b\bar b)(1-q^2\bar bb)
|j\rangle \\
&=& - {\cal C}  q^7\left\langle\matrix{6\cr 3\cr}\right\rangle_1^{1/2} q^{6n_i}|\beta|^6
(1-q^{2(n_i-1)}|\beta|^2)(1-q^{2n_i}|\beta|^2) \\
& &\times (1-q^{2(n_i+1)}|\beta|^2)\delta(i,j) \\
\end{array}
\ee
\no or
\be
m\left[\left(\frac{2}{3}~i\right)\to\left(-\frac{1}{3}~i\right)\right] = 
- {\cal C}  q^7
\left\langle\matrix{6\cr 3\cr}\right\rangle_1^{1/2}q^{6n_i}|\beta|^6f(n_i-1)f(n_i)
f(n_i+1)
\ee
\no where $f(n)$ is defined by (9.5).

This matrix element covers the following cases:
\be
u+W^-\to d \qquad \qquad c+W^-\to s \qquad \qquad t + W^-\to b
\ee
\no In particular we have by (10.17) with $n_i=0$
\be
m(u\to d) = -{\cal C} q^7\left\langle\matrix{6\cr 3\cr}\right\rangle_1^{1/2}|\beta|^6~f(-1)
f(0) f(1)
\ee
\no The ratio of matrix elements for the two reactions (10.4) and (10.17) is
\be
\frac{m\left[Q\left(-\frac{1}{3}~i\right) + W^+ \to Q\left(\frac{2}{3}~i\right)\right]}
{m\left[Q\left(\frac{2}{3}~i\right) + W^-\to Q\left(-\frac{1}{3}~i\right)\right]}
= -q_1^3
\ee
\no again up to the factor $C_+(q,\beta)/C_-(q,\beta)$.
Note that
the $-q_1^3$ appearing in (10.20) stems from the same factor in $D^3_{-30}$ that
appears in (8.5).  If the symbol for $W^+$ is defined without this factor, then the
symbols for $W^+$ and $W^-$ are $q$-conjugate and $-q_1^3$ does not appear in (10.20).
This option is subsumed in the choice of the normalizing factors
$C_+(q,\beta)$ and $C_-(q,\beta)$.  

\section{The Kobayashi-Maskawa Matrix.}

We want to compare the ratios calculated here with the Kobayashi-Maskawa
matrix, namely:
\be
\begin{array}{c|ccc}
& d & s & b \\
\hline
u & 0.973 & 0.23 & 0 \\
c & 0.24 & 0.91 & 0.06 \\
t & 0 & 0 & 1 \\
\end{array}
\ee
\no without introducing the Cabibbo-GIM angles.  (The matrix (11.1) is
known more accurately but (11.1) is adequate for the present.)

The diagonal elements are all approximately unity.  Then $q\cong 1$ in
(10.10) if $W^+$ is represented by $D^3_{-30}$ as in (7.7a).  This choice
of $W^+$, however, forbids
\be
s + W^+ \to u
\ee
\be
d + W^+ \to c
\ee
\no To include these forbidden processes as well we may replace
$D^3_{-30}$ and $D^3_{30}$ by
\be
W^+ \sim D^3_{-30}(1+a+\bar a)
\ee
\be
W^-\sim (1+a+\bar a)D^3_{30}
\ee

The expressions appearing in (11.4) and (11.5) represent minimal
modifications of $W^+$ and $W^-$.  We also have
\be
\begin{array}{rcl}
a &=& D^{1/2}_{\frac{1}{2}\frac{1}{2}} \\
\bar a &=& D^{1/2}_{-\frac{1}{2}-\frac{1}{2}} \\
\end{array}
\ee
\no so that these modified forms may be written as
\be
\begin{array}{rcl}
W^+ &\cong& D^3_{-30}(1+D^{1/2}_{\frac{1}{2}\frac{1}{2}} +
D^{1/2}_{-\frac{1}{2}-\frac{1}{2}}) \\
W^- &\cong& (1+D^{1/2}_{\frac{1}{2}\frac{1}{2}} + D^{1/2}_{-\frac{1}{2}-\frac{1}{2}})D^3_{30} \\
\end{array}
\ee
\no $W^+$ and $W^-$ may then be written as a linear combination of
$D^j_{mn}$ terms with $q$-Clebsch-Gordan coefficients.

If the factor $-q^3_1$ is dropped, $\bar D^3_{30} = D^3_{30}$ and
$\bar W^- = W^+$ in (11.4) and (11.5).

The assumptions (11.4) and (11.5) still forbid
\be
\begin{array}{rcl}
u+W^- &\to& b \\
t+W^- &\to& d \\
\end{array}
\ee
\no as required by the approximate Kobayashi-Maskawa matrix.

We would expect the justification for modifying (7.8) by (11.4) and (11.5)
for quarks to be found only in a refinement of the simple knot model 
described here.  A similar modification of (7.8) for the lepton-neutrino
system is forbidden by lepton conservation.  

According to (11.4) the matrix
element for the process:  $s+W^+\to u$ is
\be
\begin{array}{rcl}
m(s+W^+\to u) &=& \langle 0|\bar D^{3/2}_{-\frac{3}{2}-\frac{1}{2}}\cdot
D^3_{-30}(a+\bar a)D^{3/2}_{\frac{3}{2}-\frac{1}{2}}|1\rangle \\
&=& -{\cal C} \langle 0|(-a^2b)\cdot q_1^4\left\langle\matrix{6\cr 3\cr}\right
\rangle_1^{1/2}\bar b^3\bar a^3(a+\bar a)\cdot(ab^2)|1\rangle \\
&=&q_1^4 {\cal C} \left\langle\matrix{6\cr 3\cr}\right\rangle_1^{1/2}
\langle 0|(a^2b)(\bar b^3\bar a^3)a(ab^2)|1\rangle \\
\end{array}
\ee
\no where
\be
\begin{array}{rcl}
\langle 0|(a^2b)(\bar b^3\bar a^3)a(ab^2)|1\rangle &=&
\langle 0|a^2(b\bar b^3)\bar a(\bar a^2a^2)b^2|1\rangle \\
&=& q^8\langle 0|(b\bar b^3)a(a\bar a)(\bar a^2a^2)b^2|1\rangle \\
&=& q^8\langle 0|(b\bar b^3)a(1-b\bar b)(1-q^2_1b\bar b)
(1-q_1^4b\bar b)b^2|1\rangle \\
&=& q^{10}\langle 0|(b\bar b)^3(1-q^2b\bar b)(1-b\bar b)
(1-q_1^2b\bar b)a|1\rangle \\
\end{array}
\ee
\no Then by (8.10)
\[
\begin{array}{rcl}
m(s+W^+\to u) &=& q^6 {\cal C} 
\left\langle\matrix{6\cr 3\cr}\right\rangle_1^{1/2}
\langle 0|(b\bar b)^3(1-q^2b\bar b)(1-b\bar b)(1-q_1^2b\bar b)
(1-|\beta|^2)^{1/2}|0\rangle \\
&=& q^6 {\cal C} \left\langle\matrix{6\cr 3\cr}\right\rangle_1^{1/2}|\beta|^6
(1-q^2|\beta|^2)(1-|\beta|^2)(1-q_1^2|\beta|^2)(1-|\beta|^2)^{1/2} \\
\end{array}
\]
\no or
\be
m(s+W^+\to u) = q^6 {\cal C} \left\langle\matrix{6\cr 3\cr}\right\rangle_1^{1/2}
|\beta|^6 f(1) f(0) f(-1)(1-|\beta|^2)^{1/2}
\ee

The corresponding matrix element for
\[
d+W^+\to c
\]
\no is
\be
\begin{array}{rcl}
m(d+W^+\to c) &=& \langle 1|\bar D^{3/2}_{-\frac{3}{2}-\frac{1}{2}}
D^3_{30}(a+\bar a) D^{3/2}_{\frac{3}{2}-\frac{1}{2}}|0\rangle \\
&=& {\cal C} \langle 1|(-q_{1} a^2b)(-q_1^3\left\langle\matrix{6\cr 3\cr}\right\rangle_1^{1/2} \bar b^3\bar a^3)(a+\bar a)(ab^2)|0\rangle \\
&=&q_1^4 {\cal C} \left\langle\matrix{6\cr 3\cr}\right\rangle_1^{1/2}
\langle 1|a^2b(\bar b^3\bar a^3)\bar a(a\bar b)|0\rangle
\end{array}
\ee
\no Here
\be
\begin{array}{rcl}
\langle 1|(a^2b)(\bar b^3\bar a^3\bar a)ab^2|0\rangle
&=& q^8\langle 1|(b\bar b^3)a^2\bar a^3\bar aab^2|0\rangle \\
&=& q^8\langle 1|(b\bar b^3)(a^2\bar a^2)\bar a(\bar aa)b^2|0\rangle \\
&=& q^8\langle 1|(b\bar b^3)(1-b\bar b)(1-q^2b\bar b)\bar a
(1-q_1^2\bar bb)b^2|0\rangle \\
&=& q^8\langle 1|(b\bar b^3)(1-b\bar b)(1-q^2b\bar b)(1-q_1^4\bar bb)
\bar ab^2|0\rangle \\
&=& q^6\langle 1|(b\bar b)^3(1-b\bar b)(1-q^2b\bar b)(1-q_1^4\bar bb)
\bar a|0\rangle \\
&=& q^6\langle 1|q^6|\beta|^6(1-q^2|\beta|^2)(1-q^4|\beta|^2)
(1-q_1^2|\beta|^2)(1-|\beta|^2)^{1/2}|1\rangle \\
\end{array}
\ee
\no by (8.1) and (8.4).  Then by (9.5)
\be
m(d+W^+\to c) = q^8 {\cal C} \left\langle\matrix{6\cr 3\cr}\right\rangle_1^{1/2}
|\beta|^6 f(1)f(2) f(-1)(1-|\beta|^2)^{1/2}
\ee
\no By (11.11) and the preceding equation
\be
\begin{array}{rcl}
\frac{m(s+W^+\to u)}{m(d+W^+\to c)} &=& \frac{q^6}{q^8}
\frac{f(1) f(0) f(-1)}{f(1) f(2) f(-1)}\\
&=& q^{-2}\frac{f(0)}{f(2)} = q^{-2}
\frac{1-|\beta|^2}{1-q^4|\beta|^2}
\end{array}
\ee
\no Since the corresponding ratio in the Kobayashi-Maskawa matrix is very
close to unity, Eq. (11.15) again implies
\be
q\cong 1
\ee
\no One also finds by (10.4) and (11.14)
\be
\frac{m(d+W^+\to u)}{m(d+W^+\to c)} = q_1^4 \frac{f(0)}{f(2)}
\frac{1}{(1-|\beta|^2)^{1/2}}
\ee
\no By (11.15)
\be
\begin{array}{rcl}
\frac{m(d+W^+\to u)}{m(d+W^+\to c)} &=& q_1^2
\frac{m(s+W^+\to u)}{m(d+W^+\to c)} \frac{1}{(1-|\beta|^2)^{1/2}} \\
&\cong& \frac{1}{(1-|\beta|^2)^{1/2}}
\end{array}
\ee
\no if we set $q\cong 1$ according to (11.16). 

From the Kobayashi-Maskawa
matrix we have
\be
\frac{m(d+W^+\to u)}{m(d+W^+\to c)} = 4.054
\ee
\no By (11.18) and (11.19)
\be
|\beta| \cong .968
\ee
\no Again $q$ and $|\beta|$ are approximately unity.

We next compare with the small Kobayashi-Maskawa entry $(cb)$
\be
\begin{array}{rcl}
m(b+W^+\to c) &=& \langle 1|\bar D^{3/2}_{-\frac{3}{2}-\frac{1}{2}}\cdot
D^3_{-30}(a+\bar a)D^{3/2}_{\frac{3}{2}-\frac{1}{2}}|2\rangle \\
&=& {\cal C}\langle 1|(-q_{1}a^2b)(-q_1^3\left\langle\matrix{6\cr 3\cr}
\right\rangle_1^{1/2}\bar b^3\bar a^3)(a+\bar a)(ab^2)|2\rangle \\
&=& q_1^4 {\cal C} \left\langle\matrix{6\cr 3\cr}\right\rangle_1^{1/2}
\langle 1|a^2b\bar b^3\bar a^3(a+\bar a)ab^2|2\rangle \\
\end{array}
\ee
\no where (11.21) may be reduced as follows:
\be
\begin{array}{rcl}
\langle 1|a^2b\bar b^3\bar a^3(a+\bar a)ab^2|2\rangle &=&
q^8\langle 1|b\bar b^3a^2\bar a^3a^2b^2|2\rangle \\
&=& q^{10}\langle 1|(b\bar b)^3(a^2\bar a^2)(\bar aa)a|2\rangle \\
&=& q^{16}|\beta|^6(1-q^2|\beta|^2)(1-q^4|\beta|^2)
(1-|\beta|^2)(1-q^2|\beta|^2)^{1/2} \\
\end{array}
\ee
\no Then
\be
m(b+W^+\to c) = q^{12} {\cal C} \left\langle
\matrix{6\cr 3\cr}\right\rangle_1^{1/2}|\beta|^6
(1-q^4|\beta|^2)(1-|\beta|^2)(1-q^2|\beta|^2)^{3/2}
\ee
\no By (10.4) and the preceding equation
\be
\frac{m(b+W^+\to c)}{m(d+W^+\to u)} = q^8
\frac{1-q^4|\beta|^2}{1-q_1^2|\beta|^2} (1-q^2|\beta|^2)^{1/2}
\ee
\no By comparing with the Kobayashi-Maskawa matrix one has
\be
q^8\frac{1-q^4|\beta|^2}{1-q_1^2|\beta|^2}
(1-q^2|\beta|^2)^{1/2} = .0617
\ee

If one sets $q=1$ as in the previous case, one finds:
\be
\begin{array}{rcl}
(1-|\beta|^2)^{1/2} &=& .0617 \\
|\beta| &=& .998 \\
\end{array}
\ee
\no The approximate solution for this case is then $(q,\beta)
\cong (1.00,.998)$.  

Finally the $(ts)$ element according to our model is
\be
m(s+W^+\to t) = \langle 2|\bar D^{3/2}_{-\frac{3}{2}-\frac{1}{2}}\cdot
D^3_{-30}(a+\bar a)\cdot D^{3/2}_{\frac{3}{2}-\frac{1}{2}}|1\rangle
\ee
\no or
\be
m(s+W^+\to t) = q_1^4 {\cal C} 
\left\langle\matrix{6\cr 3\cr}\right\rangle_1^{1/2}
\langle 2|a^2b\bar b^3\bar a^3\bar a(-ab^2)|1\rangle
\ee
\no We find
\be
\begin{array}{rcl}
\frac{m(s+W^+\to t)}{m(b+W^+\to c)} &=& q^2
\frac{(1-q^6|\beta|^2)(1-q^4|\beta|^2)(1-|\beta|^2)
(1-q^2|\beta|^2)^{1/2}}{(1-q^4|\beta|^2)(1-|\beta|^2)(1-q^2|\beta|^2)^
{3/2}} \\
&=& q^2 \frac{1-q^6|\beta|^2}{1-q^2|\beta|^2} \\
\end{array}
\ee
\no If $q=1$, the matrix elements, $(ts)$ and $(cb)$, are equal.
The vanishing $(ts)$ entry in the KM matrix may be compatible with
(11.29) and the small $(cb)$ value already computed.

The $(ub)$ and $(dt)$ matrix elements vanish for (11.4) and (11.5) and
also in the approximate KM matrix (11.1).

We have ignored the phase factors appearing in the empirical matrix
elements as well as the phase factors stemming from $\lambda_n$
and $\mu_n$ (Eqs. (8.10) and (8.11)) and therefore appearing in the
computed matrix elements as well.  The results of this 
section are summarized in Table 3.
\ve

\no {\it Results of Comparing with K.M. Matrix (without Cabibbo-GIM Angles).}

\[
\begin{array}{ccccc}
\underline{\rm Ratio~of~Matrix~Elements} & \underline{\rm Ratio~from~ Model} & \underline{\rm K.M.~ Matrix} & \qquad\underline{q} &
\qquad\underline{\beta} \\
\frac{m(s+W^+\to u)}{m(d+W^+\to c)} & 
q^{-2}\left(\frac{1-|\beta|^2}{1-q^4|\beta|^2}\right) & 
\frac{.23}{.24} = .958 & \qquad\sim 1 & \qquad\sim 1 \\
& & & & \\
\frac{m(d+W^+\to u)}{m(d+W^+\to c)} &
\frac{m(s+W^+\to u)}{m(d+W^+\to c)}\frac{q^{-2}}{(1-|\beta|^2)^{1/2}} &  &  & \\
& = .958 \frac{q^{-2}}{(1-|\beta|^2)^{1/2}} & \frac{.973}{.24} = 4.05 &
\qquad\sim 1 & \qquad \sim .968 \\
& & & & \\
\frac{m(b+W^+\to c)}{m(d+W^+\to u)} &
q^8\frac{1-q^4|\beta|^2}{1-q^{-2}|\beta|^2}(1-q^2|\beta|^2)^{1/2} &
\frac{.06}{.973} = .0617 & \qquad\sim 1 & \qquad \sim .998 \\
& & & & \\
\frac{m(s+W^+\to t)}{m(b+W^+\to c)} &
q^2\frac{1-q^6|\beta|^2}{1-q^2|\beta|^2} & \sim 1 & \qquad\sim 1 & \qquad\sim 1 \\
\end{array}
\] 
\begin{center}
{\bf Table 3.}
\end{center}

\no Compare with $(q=1,|\beta| = .707)$ for lepton-neutrino production.

In comparing with the KM matrix we have found that $(q,\beta)$ remain
stable and close to unity.

\section{Neutral Couplings.}

\no {\underline{Lepton-Lepton Interactions.}}
\vskip.3cm

\no (a) Mediated by $W_3:$

The matrix element for
\[
\ell(n) + W_3 \to \ell^\prime(n)
\]
\no is by (8.2), (8.3), and (8.4)
\be
\begin{array}{rcl}
\langle n|\bar D^{3/2}_{\frac{3}{2}\frac{3}{2}}D^3_{00}
D^{3/2}_{\frac{3}{2}\frac{3}{2}}|n\rangle &=&
\langle n|\bar a^3 f_3(b\bar b) a^3|n\rangle \\
&=& \langle n|\bar a^3a^3|n\rangle f_3(q^{-6+2n}|\beta|^2) \\
\end{array}
\ee

\vskip.3cm

\no (b) Mediated by $W_0:$

The corresponding matrix element for
\[
\ell(n) + W_0 \to \ell^\prime(n)
\]
\no is by (8.2), (8.3), and (8.5)
\be
\begin{array}{rcl}
\langle n|\bar D^{3/2}_{\frac{3}{2}\frac{3}{2}} D^3_{-11}
D^{3/2}_{\frac{3}{2}\frac{3}{2}}|n\rangle &=&
\langle n|\bar a^3 \hat f_0(b\bar b)\bar b^2 a^3|n\rangle \\
&=& q^{-6}\langle n|\bar a^3a^3|n\rangle
\hat f_0(q^{-6+2n}|\beta|^2)q^{2n}\bar\beta^2 \\
\end{array}
\ee
\no where by (8.6)
\[
f_0(b,\bar b) = \hat f_0(b\bar b)\bar b^2
\]
\no Then the ratio of the $W_3$ to the $W_0$ matrix elements is 
\be
\frac{\langle n|\bar D^{3/2}_{\frac{3}{2}\frac{3}{2}}
D^3_{00}D^{3/2}_{\frac{3}{2}\frac{3}{2}}|n\rangle}
{\langle n|\bar D^{3/2}_{\frac{3}{2}\frac{3}{2}}D^3_{-11}
D^{3/2}_{\frac{3}{2}\frac{3}{2}}|n\rangle} = 
\frac{f_3(q^{-6+2n}|\beta|^2)q_1^{2n-6}}{\hat f_0(q^{-6+2n}|\beta|^2)
\bar\beta^2}
\ee

Since we carry over the Weinberg-Salam relation between
$(W^3,W^0)$ and $(A,Z)$, we have
\be
\langle n|A|n\rangle = \langle n|\bar a^3a^3|n\rangle
[\bar\beta^2q^{-6}C_0\hat f_0(q^{-6+2n}|\beta|^2)
\cos\theta + C_3f_3(q^{-6+2n}|\beta|^2)\sin\theta] 
\ee
\be
\langle n|Z|n\rangle = \langle n|\bar a^3a^3|n\rangle
[-\bar\beta^2q^{-6}C_0\hat f_0(q^{-6+2n}|\beta|^2)\sin\theta +C_3 f_3
(q^{-6+2n}|\beta|^2)\cos\theta] \\
\ee
\vskip.3cm

\no \underline{Neutrino-Neutrino Interactions}.

\no (a) Mediated by $W_3$.
\be
\begin{array}{rcl}
\langle n|\bar D^{3/2}_{-\frac{3}{2}\frac{3}{2}}D^3_{00}
D^{3/2}_{-\frac{3}{2}\frac{3}{2}}|n\rangle &=&
q_1^6\langle n|b^3f_3(b\bar b)\bar b^3|n\rangle \\
&=& q^{6n-6}|\beta|^6f_3(q^{2n}|\beta|^2) 
\end{array}
\ee

\no (b) Mediated by $W_0$.
\be
\langle n|\bar D^{3/2}_{-\frac{3}{2}\frac{3}{2}} D^3_{-11}
D^{3/2}_{-\frac{3}{2}\frac{3}{2}}|n\rangle =
q^{8n-6}|\beta|^6\hat f_0(q^{2n}|\beta|^2)\bar\beta^2
\ee
\no The ratio of these matrix elements is
\be
\frac{\langle n|\bar D^{3/2}_{-\frac{3}{2}\frac{3}{2}}D^3_{00}
D^{3/2}_{-\frac{3}{2}\frac{3}{2}}|n\rangle}
{\langle n|\bar D^{3/2}_{-\frac{3}{2}\frac{3}{2}}
D^3_{-11}D^{3/2}_{-\frac{3}{2}\frac{3}{2}}|n\rangle} = q^{-2n}
\frac{f_3(q^{2n}|\beta|^2)}{\hat f_0(q^{2n}|\beta|^2)}
\frac{1}{\bar\beta^2}
\ee
\no Let
\begin{eqnarray}
\hat f_0^\prime &=& C_0\hat f_0  \nonumber \\
f_3^\prime &=& C_3f_3 
\end{eqnarray}
\no The matrix elements for $A$ and $Z$ are
\be
\langle n|A|n\rangle = q^{6n-6}|\beta|^6
[q^{2n}\hat f_0^\prime(q^{2n}|\beta|^2)\bar\beta^2\cos\theta +
f_3^\prime(q^{2n}|\beta|^2)\sin\theta]
\ee
\be
\langle n|Z|n\rangle = q^{6n-6}|\beta|^6[-q^{2n}\hat f_0^\prime(q^{2n}|\beta|^2)
\bar\beta^2\sin\theta + f_3^\prime(q^{2n}|\beta|^2)\cos\theta]
\ee
\no Since $\langle n|A|n\rangle$ must vanish for neutrinos one demands
\be
q^{2n}\hat f_0^\prime(q^{2n}|\beta|^2)
\bar\beta^2\cos\theta + f_3^\prime(q^{2n}|\beta|^2)\sin\theta = 0
\ee
\no or
\be
\tan\theta = -q^{2n}\left(\frac{\hat f_0^\prime(q^{2n}|\beta|^2)}
{f_3^\prime(q^{2n}|\beta|^2)}\right)\bar\beta^2
\ee
\no For the neutrino family we set $\beta = i|\beta|$, since we take
$\tan\theta$ positive.

The requirement (12.12) is equivalent to the requirement of standard 
theory that the photon interacts only with electric charge and not at all with hypercharge.  Eq. (12.13) states that $\tan\theta$ is so chosen
that $W_3$ and $W_0$ are mixed so that the photon has no role in the
weak interactions.

Then
\be
q^{2n} \frac{\hat f_0^\prime(q^{2n}|\beta|^2)|\beta|^2}
{f_3^\prime(q^{2n}|\beta|^2)} = \tan\theta
\ee
\no One requires that the Weinberg angle be independent of $n$.  Then
$q=1$ and
\be
\begin{array}{rcl}
\frac{C_0}{C_3}\frac{\hat f_0(|\beta|^2)|\beta|^2}
{f_3(|\beta|^2)} &=& \tan\theta \\
&=& g^\prime/g \\
&=& .528 \\
\end{array}
\ee

\ve

\section {Charge-Retention Interactions for Quarks.}

We consider first the couplings of the $(dsb)$ family to
$W_3$ and $W_0$.

\no (a) Mediated by $W_3$:
\be
Q(n) + W_3 \to Q(n)^\prime
\ee

The matrix element for (13.1) is
\be
\begin{array}{rcl}
\langle n|\bar D^{3/2}_{\frac{3}{2}-\frac{1}{2}} D^3_{00}
D^{3/2}_{\frac{3}{2}-\frac{1}{2}}|n\rangle &=&
{\cal C}' \langle n|(\bar b^2\bar a)\cdot f_3(b\bar b)\cdot (ab^2)|n\rangle \\
&=& {\cal C}'  \langle n|\bar b^2(\bar aa)f_3(q_1^2b\bar b)b^2|n\rangle\\
&=& {\cal C}'\langle n|(\bar bb)^2(\bar aa)|n\rangle
\langle n|f_3(q_1^2(b\bar b)|n\rangle \\
\end{array}
\ee
\no with
\[ {\cal C}' = \left( \Delta^{3/2}_{\frac{3}{2}-\frac{1}{2}}   
\left\langle\matrix{3\cr 1\cr}\right\rangle_1 \right)^{2} \]

\no (b) Mediated by $W_0$:
\be
Q(n) + W_0 \to Q(n)^\prime
\ee
\no with the following matrix element:
\be
\begin{array}{rcl}
\langle n|\bar D^{3/2}_{\frac{3}{2}-\frac{1}{2}}\cdot D^3_{-11}\cdot
D^{3/2}_{\frac{3}{2}-\frac{1}{2}}|n\rangle &=& {\cal C}'
\langle n|\bar b^2\bar a\cdot\hat f_0(b\bar b)\bar b^2\cdot ab^2
|n\rangle \\
&=& {\cal C}' q_1^2\langle n|(\bar bb)^2(\bar aa)|n\rangle
\langle n|\hat f_0(q_1^2\bar bb)\bar b^2|n\rangle \\
\end{array}
\ee
\no Then the ratio of the $W_3$ to the $W_0$ matrix elements is
\be
\frac{\langle n|\bar D^{3/2}_{\frac{3}{2}-\frac{1}{2}}
D^3_{00}D^{3/2}_{\frac{3}{2}-\frac{1}{2}}|n\rangle}
{\langle n|\bar D^{3/2}_{\frac{3}{2}-\frac{1}{2}} D^3_{-11}
D^{3/2}_{\frac{3}{2}-\frac{1}{2}}|n\rangle} = q^2
\frac{\langle n|f_3(q_1^2b\bar b)|n\rangle}
{\langle n|\hat f_0(q_1^2b\bar b)\bar b^2|n\rangle}
\ee
\no and we also have
\be
\langle n|A|n\rangle = \langle n|(\bar bb)^2\bar aa|n\rangle
[q_1^2\langle n|\hat f_0^\prime(q_1^2\bar bb)\bar b^2|n\rangle
\cos\theta + \langle n|f_3^\prime(q_1^2\bar bb)|n\rangle \sin\theta]
\ee
\be
\langle n|Z|n\rangle = \langle n|(\bar bb)^2\bar aa|n\rangle
[-q_1^2\langle n|\hat f_0^\prime(q_1^2\bar bb)\bar b^2|n\rangle
\sin\theta + \langle n|f_3^\prime(q_1^2\bar bb)|n\rangle\cos\theta]
\ee
\no We next consider the corresponding couplings of the $uct$-family
\vskip.3cm

\no (a) Mediated by $W_3$:
\vskip.3cm

The matrix elements are
\be
\begin{array}{rcl}
\langle n|\bar D^{3/2}_{-\frac{3}{2}-\frac{1}{2}}D^3_{00}
D^{3/2}_{-\frac{3}{2}-\frac{1}{2}}|n\rangle &=& {\cal C}' q_{1}^2
\langle n|a^2b\cdot f_3(b\bar b)\cdot\bar b\bar a^2|n\rangle \\
&=& {\cal C}' q_{1}^2 q^4\langle n|(b\bar b)(a^2\bar a^2)|n\rangle
\langle n|f_3(q^4b\bar b)|n\rangle \\
\end{array}
\ee

\no (b) Mediated by $W_0$:
\be
\begin{array}{rcl}
\langle n|\bar D^{3/2}_{-\frac{3}{2}-\frac{1}{2}}D^3_{-11}
D^{3/2}_{-\frac{3}{2}-\frac{1}{2}}|n\rangle &=& {\cal C}' q_{1}^2
\langle n|(a^2b)\cdot(\hat{f}_0(b\bar b)\bar b^2)
\cdot\bar b\bar a^2|n\rangle \\
&=& {\cal C}' q_{1}^2 q^8\langle n|(b\bar b)(a^2\bar a^2)|n\rangle
\langle n|\hat f_0(q^4b\bar b)\bar b^2|n\rangle \\
\end{array}
\ee

Then the ratio of the $W_3$ to the $W_0$ matrix elements is
\be
\frac{\langle n|\bar D^{3/2}_{-\frac{3}{2}-\frac{1}{2}}D^3_{00}
D^{3/2}_{-\frac{3}{2}-\frac{1}{2}}|n\rangle}
{\langle n|\bar D^{3/2}_{-\frac{3}{2}-\frac{1}{2}}D^3_{-11}
D^{3/2}_{-\frac{3}{2}-\frac{1}{2}}|n\rangle} =
q^{-4}\frac{\langle n|f_3(q^4b\bar b)|n\rangle}
{\langle n|\hat f_0(q^4b\bar b)\bar b^2|n\rangle}
\ee
\no and the $A$ and $Z$ matrix elements are
\be
\langle n|A|n\rangle = \langle n|(b\bar b)(a^2\bar a^2)|n\rangle
[q^8\langle n|\hat f_0^\prime(q^4b\bar b)\bar b^2|n\rangle\cos\theta +
q^4\langle n|f_3^\prime(q^4\bar bb)|n\rangle\sin\theta]
\ee
\be
\langle n|Z|n\rangle = \langle n|(b\bar b)(a^2\bar a^2)|n\rangle
[-q^8\langle n|\hat f_0^\prime(q^4b\bar b)\bar b^2|n\rangle
\sin\theta+q^4\langle n|f_3^\prime(q^4\bar bb)|n\rangle\cos\theta]
\ee

\vskip.5cm

\section{Decays of the $Z^0$.}

\no \underline{Decays of the $Z^0$ into Leptons}.

The rates of these decays are described by
\[
\Gamma_n(Z^0 \to \ell + \bar\ell) \sim
\left|\langle n|\bar D^{3/2}_{\frac{3}{2}\frac{3}{2}}
(C_3D^3_{00}\cos\theta - C_0D^3_{-11}\sin\theta)
D^{3/2}_{\frac{3}{2}\frac{3}{2}}|n\rangle\right|^2 
\]
\no or
\[
\hfil ~~~~ = |a_n\cos\theta - b_n\sin\theta|^2
\]
\no where
\be
a_n = C_3\langle n|\bar D^{3/2}_{\frac{3}{2}\frac{3}{2}}
D^3_{00}D^{3/2}_{\frac{3}{2}\frac{3}{2}}|n\rangle
\ee
\be
b_n = C_0\langle n|\bar D^{3/2}_{\frac{3}{2}\frac{3}{2}}D^3_{-11}
D^{3/2}_{\frac{3}{2}\frac{3}{2}}|n\rangle
\ee
\no Then
\be
\begin{array}{rcl}
\frac{\Gamma_{n+1}}{\Gamma_n} &=&
\left|\frac{a_{n+1}\cos\theta-b_{n+1}\sin\theta}
{a_n\cos\theta-b_n\sin\theta}\right|^2 \\
& & \\
&=& \left|\frac{1-\frac{b_{n+1}}{a_{n+1}}\tan\theta}
{1-\frac{b_n}{a_n}\tan\theta}\right|^2
\left|\frac{a_{n+1}}{a_n}\right|^2 \\
\end{array}
\ee
\no One finds
\be
a_n = \langle n|\bar a^3a^3|n\rangle f_3^\prime(q^{-6+2n}|\beta|^2)
\ee
\be
b_n = q^{-6}\langle n|\bar a^3a^3|n\rangle
\hat f_0^\prime(q^{-6+2n}|\beta|^2)\bar\beta^2
\ee
\no Then
\be
\frac{b_n}{a_n} = q^{-6} 
\frac{\hat f_0^\prime(q^{-6+2n}|\beta|^2)\bar\beta^2}
{f_3^\prime(q^{-6+2n}|\beta|^2)}
\ee
\be
\frac{a_{n+1}}{a_n} =
\frac{f_3^\prime(q^{-4+2n}|\beta|^2)}
{f_3^\prime(q^{-6+2n}|\beta|^2)}~
\frac{\langle n+1|\bar a^3a^3|n+1\rangle}
{\langle n|\bar a^3a^3|n\rangle}
\ee
\no If $q=1$, then by (14.7)
\be
\left|\frac{a_{n+1}}{a_n}\right|^2 = 1
\ee
\no If $q=1$, then by (14.6)
\be
\frac{b_{n+1}}{a_{n+1}} = \frac{b_n}{a_n}
\ee
\no and by (14.3)
\be
\frac{\Gamma_{n+1}}{\Gamma_n} = 1
\ee
\no The measured rates are$^7$
\be
\frac{\Gamma(\mu^+\mu^-)}{\Gamma(e^+e^-)} = 1.0009
\ee
\be
\frac{\Gamma(\tau^+\tau^-)}{\Gamma(e^+e^-)} = 1.0019
\ee
\no The measured rates are thus compatible with $q\cong 1$.

\vskip.3cm

\no \underline{Decay of $Z^0$ into Neutrinos}.

We now have
\be
\frac{\Gamma(Z^0\to\nu_e+\bar\nu_e)}
{\Gamma(Z^0\to e+\bar e)} =
\left|\frac{\langle n|\bar D^{3/2}_{-\frac{3}{2}\frac{3}{2}}
(C_3D^3_{00}\cos\theta-C_0D^3_{-11}\sin\theta)
D^{3/2}_{-\frac{3}{2}\frac{3}{2}}|n\rangle}
{\langle n|\bar D^{3/2}_{\frac{3}{2}\frac{3}{2}}
(C_3D^3_{00}\cos\theta-C_0D^3_{-11}\sin\theta)
D^{3/2}_{\frac{3}{2}\frac{3}{2}}|n\rangle}\right|^2
\ee

Let
\be
R = \frac{\Gamma(Z^0\to\nu_e+\bar\nu_e)}
{\Gamma(Z^0\to e+\bar e)}
\ee
\no Then
\be
R = \left|\frac{|\beta|^6}{(1-|\beta|^2)^3}\right|^2 \qquad \mbox{if} ~~ q=1
\ee
\no The measured value of $R$ is 1.98.$^7$
\no Then
\be
\frac{|\beta|^2}{1-|\beta|^2} = (1.98)^{1/6}
\ee
\no and
\be
|\beta| = .727
\ee
\no Compare $(q,\beta)$ = (1,.727) with $(q,\beta)$ = (1,.707), the
values found earlier in the discussion of lepton-neutrino production by charged $W$ (with the assumption of the universal Fermi
interaction).

\vskip.3cm

\no \underline{Pair Production of (2/3) Quarks}.

In this case
\be
\begin{array}{rcl}
Z^0 &\to& u+\bar u \\ &\to& c+\bar c \\ &\to& t+\bar t \\
\end{array}
\ee

The relevant matrix element is
\be
\langle n|\bar D^{3/2}_{-\frac{3}{2}-\frac{1}{2}}
(C_3D^3_{00}\cos\theta-C_0D^3_{-11}\sin\theta)
D^{3/2}_{-\frac{3}{2}-\frac{1}{2}}|n\rangle
\ee
\no so that
\be
\Gamma_n\sim|\langle n|a^2\bar a^2|n\rangle q^{2n-2}\langle 3\rangle_1
|\beta|^2[(f_3^\prime(q^{2n+4}|\beta|^2)\cos\theta-f_0^\prime
(q^{2n+4}|\beta|^2)\sin\theta|^2
\ee
\no and
\be
\frac{\Gamma_{n+1}}{\Gamma_n} =
\left|\frac{\langle n+1|a^2\bar a^2|n+1\rangle}
{\langle n|a^2\bar a^2|n\rangle} q^2
\frac{[f_3^\prime(q^{2n+6}|\beta|^2)\cos\theta-f_0^\prime(q^{2n+6}|\beta|^2)
\sin\theta]}
{f_3^\prime(q^{2n+4}|\beta|^2)\cos\theta-f_0^\prime(q^{2n+4}|\beta|^2)\sin\theta]}\right|^2
\ee

If $q\cong 1$, there is little dependence of this ratio on $n$ and
the three rates (14.22) are approximately equal.

\vskip.3cm

\no \underline{Pair Production of (-1/3) Quarks}.

Now
\be
\begin{array}{rcl}
Z^0 &\to& d+\bar d \\ &\to& s+\bar s \\ &\to& b+\bar b \\
\end{array}
\ee
\no with the matrix element
\be
\langle n|\bar D^{3/2}_{\frac{3}{2}-\frac{1}{2}}
[C_3D^3_{00}\cos\theta-C_0D^3_{-11}\sin\theta]
D^{3/2}_{\frac{3}{2}-\frac{1}{2}}|n\rangle
\ee
\no so that
\be
\Gamma_n \sim |q^{4n}|\beta|^4\langle 3\rangle_1
\langle n|\bar aa|n\rangle[f_3^\prime(q^{2n-2}|\beta|^2)
\cos\theta-f_0^\prime(q^{2n-2}|\beta|^2)\sin\theta]|^2
\ee
\no and
\be
\frac{\Gamma_{n+1}}{\Gamma_n} = \left|
\frac{\langle n+1|\bar aa|n+1\rangle}{\langle n|a\bar a|n\rangle}
\frac{f_3^\prime(q^{2n}|\beta|^2)\cos\theta-f_0^\prime(q^{2n}|\beta|^2)\sin\theta}
{f_3^\prime(q^{2n-2}|\beta|^2)\cos\theta-f_0^\prime(q^{2n-2}|\beta|^2)\sin\theta}\right|^2
\ee

If $q\cong 1$, the three rates (14.22) are again approximately equal.

\vskip.3cm

\no \underline{Ratio of Pair Production of 2/3 and -1/3 Quarks by $Z^0$}.

Let
\be
R = \frac{\Gamma\left(Z^0\to Q\left(-\frac{1}{3}\right) + 
\bar Q\left(-\frac{1}{3}\right)\right)}
{\Gamma\left(Z^0\to Q\left(\frac{2}{3}\right) +
\bar Q\left(\frac{2}{3}\right)\right)}
\ee
\no By (13.7) and (13.12)
\be
R = \left|q^{2n+2}|\beta|^2\frac{\langle n|\bar aa|n\rangle}
{\langle n|a^2\bar a^2|n\rangle}
\frac{[f_3^\prime(q^{2n-2}|\beta|^2)\cos\theta-f_0^\prime(q^{2n-2}|\beta|^2)\sin\theta]}
{[f_3^\prime(q^{2n+4}|\beta|^2)\cos\theta-f_0^\prime(q^{2n+4}|\beta|^2)\sin\theta]}\right|^2
\ee
\no If $q\cong 1$,
\be
R\cong \left|\frac{|\beta|^2}{1-|\beta|^2}\right|^2
\ee

By Ref. 6, the fractions of all decays of the $Z^0$ into $(u\bar u+c\bar c)/2$ and $(d\bar d+s\bar s+b\bar b)/3$ are 10.1\% and 16.6\%
respectively.  One may then estimate $\beta$ by setting
\be
\left|\frac{|\beta|^2}{1-|\beta|^2}\right|^2 = \frac{16.6}{10.1}
\ee
\no Then
\be
|\beta| = .750
\ee
\no One may compare $(q,\beta)$ for the following examples:
\be
\begin{array}{lc}
& \underline{q,\beta} \\
\Gamma[W^+\to e^+ + \bar\nu_e] & (1,.707) \\
\Gamma[Z^0\to e+\bar e)/\Gamma(Z^0\to \nu_e+\bar\nu_e)] & (1,.727) \\
\Gamma\left[\left(Z^0\to Q\left(-\frac{1}{3}\right) +
\bar Q\left(-\frac{1}{3}\right)\right)
/\left(Z^0\to Q\left(\frac{2}{3}\right) +
\bar Q\left(\frac{2}{3}\right)\right)\right] & (1,.750) \\
\end{array}
\ee

In these tests the value of $q$ is simply assigned:  $q=1$.  Although
a closer fit is possible if $q$ is allowed to vary, one already sees
that $\beta$ is stable.

\vskip.3cm

\no \underline{Relative Production of Lepton Pairs by $A$ and $Z$}.

\vskip.3cm

By (12.4) and (12.5)
\be
\frac{\langle n|A|n\rangle}{\langle n|Z|n\rangle} =
\frac{S+\tan\theta}{-S\tan\theta+1}
\ee
\no where $|n\rangle$ is any lepton state and
\be
S = q^{-6}|\beta|^2
\frac{\hat f_0^\prime(q^{-6+2n}|\beta|^2)}{f_3^\prime(q^{-6+2n}|\beta|^2)}
\ee
\no since $\bar\beta = \beta$ for lepton states.

By (12.14)
\be
\tan\theta= q^{2n}|\beta|^2 
\frac{\hat f_0^\prime(q^{2n}|\beta|^2)}{f_3^\prime(q^{2n}|\beta|^2)}
\ee
\no and since $\tan\theta$ is independent of $n$, one has
$q=1$ and by (14.33)
\be
S = \tan\theta
\ee
\no Then by (14.32)
\be
\frac{\langle n|A|n\rangle}{\langle n|Z|n\rangle} =
\frac{2\tan\theta}{1-\tan^2\theta}
\ee
\no in agreement with the Weinberg-Salam model for the ratio
$Q/Q^\prime$ of the electric to the hypercharge.

\vskip.5cm

\section {Covariant Derivative of Neutral States.}

We are replacing the standard $SU(2)_L\times U(1)$ theory by a knot theory
based on $SU_q(2)_L$ alone, i.e. we are assuming that the roles of
charge and hypercharge in the standard theory can be carried by $SU_q(2)$
alone.

The transition from $SU(2)\times U(1)$ to $SU_q(2)$ may be partially
described as follows:
\be
g^\prime W_0t_0 + g\bar W\vec t \to \hat g[C_0W_0D^3_{-11} +
C_3W_3D^3_{00} + C_-W_-D^3_{30} + C_+W_+D^3_{-30}]
\ee
\no where the $C_\tau$ $(\tau = 0,3,-,+)$ are functions of $q$ and $\beta$.  Here $\hat g$ is the coupling constant of the $SU_q(2)$ theory while
$g$ and $g^\prime$ are the usual coupling constants of the $SU(2)\times
U(1)$ model.

The neutral couplings in the knot theory are then described by
\be
\hat g[C_0W_0D^3_{-11} + C_3W_3D^3_{00}]
\ee

By carrying over the relation between $(W_0,W_3)$ and $(A,Z)$ from
the standard theory, we replace (15.2) by
\be
\hat g[{\cal A}A + {\cal Z}Z]
\ee
\no where
\be
{\cal A} = C_0 \cos\theta D^3_{-11} + C_3
\sin\theta D^3_{00}
\ee
\be
{\cal Z} = -C_0\sin\theta D^3_{-11} + C_3\cos\theta
D^3_{00}
\ee

Since ${\cal A}$ and ${\cal Z}$ lie in the $(b,\bar b)$ subalgebra,
neutrino states are eigenstates of these operators.  Since 
${\cal A}$ is further restricted by the physical condition that
photons do not interact with neutrinos, we have
\be
\bar\nu {\cal A}\nu = \bar\nu{\cal A}^\prime\nu =
\bar\nu\nu{\cal A}^\prime = 0
\ee
\no Then by (15.4)
\be
C_0\cos\theta D_{-11}^{3~\prime} + C_3\sin\theta
D^{3~\prime}_{00} = 0
\ee
\no and by (15.5)
\begin{eqnarray}
{\cal Z}\nu &=& [-C_0\sin\theta D_{11}^{3~\prime} +
C_3\cos\theta D_{00}^{3~\prime}]\nu \\
&=& \frac{C_3}{\cos\theta} D_{00}^{3~\prime}\nu
\end{eqnarray}

The complete covariant derivative on a neutrino state, or any
neutral state, is then by (15.1), (15.3), (15.6) and (15.9)
\be
\nabla_\mu = \partial_\mu + ig\left[C_-W_-D^3_{30} +
C_+W_+D^3_{-30} + \frac{C_3}{\cos\theta} D^3_{00}\right]
\ee
\no In addition
\be
C_0 = -C_3\left(\frac{D^\prime_{00}}
{D_{-11}^{3~\prime}}\right) \tan\theta
\ee
\no by (15.7).

\vskip.5cm

\section {Kinetic Energy of Neutral Higgs Scalar and Vector Masses.}

Let us assign the neutral Higgs scalar to the lowest state of the
trefoil previously identified with the neutrino family, namely
(-3,2) carrying the representation $D^{3/2}_{-\frac{3}{2}\frac{3}{2}}$.
This trefoil lies entirely in the $(b,\bar b)$ subalgebra as is also
the case for the neutral vectors $W_0$ and $W_3$.  We then have
\be
\hat\varphi = \rho D^{3/2}_{-\frac{3}{2}\frac{3}{2}}|0\rangle
\ee
\no where $\hat\varphi$ is the neutral Higgs scalar and the covariant
derivative of this field is by (15.10)
\be
\nabla_\mu\hat\varphi = \{\partial_\mu\rho + i\hat g\rho
[C_-W_{-\mu}D^3_{30} + C_+W_{+\mu}D^3_{-30} +
\frac{C_3}{\cos\theta} Z_\mu D^3_{00}]\} D^3_{-\frac{3}{2}\frac{3}{2}}
|0\rangle
\ee
\no The kinetic energy of the lowest state of the neutral Higgs scalar is
\be
\begin{array}{rcl}
\overline{\nabla^\mu\hat\varphi}~ \nabla_\mu\hat\varphi &=&
\langle 0|\bar D^{3/2}_{-\frac{3}{2}\frac{3}{2}}
\{\partial^\mu\rho\partial_\mu\rho + \hat g^2\rho^2
[|C_-|^2 W_-^\mu W_{-\mu}\bar D^3_{30}D^3_{30} + |C_+|^2
W_+^\mu W_{+\mu}\bar D^3_{-30}D^3_{-30} \\
& & + \frac{|C_3|^2}{\cos^2\theta} Z^\mu Z_\mu\bar D^3_{00}D^3_{00}]\}
D^3_{-\frac{3}{2}\frac{3}{2}}|0\rangle
\end{array}
\ee
\no where we have used orthogonality of the $D^3_{mn}$ that follows from
\be
\langle 0|\ldots~\bar a^na^m~\ldots|0\rangle \sim \delta^{nm}
\ee
\no Then
\be
\overline{\nabla^\mu\hat\varphi}~\nabla_\mu\hat\varphi =
I~\partial^\mu\rho\partial_\mu\rho + \hat g^2\rho^2
[II|C_-|^2W_-^\mu W_{-\mu} + III|C_+|^2 W_+^\mu W_{+\mu} +
\frac{IV|C_3|^2}{\cos^2\theta} Z^\mu Z_\mu]
\ee
\no where
\be
\begin{array}{rcl}
I &=& \langle 0|\bar D^{3/2}_{-\frac{3}{2}\frac{3}{2}}
D^{3/2}_{-\frac{3}{2}\frac{3}{2}}|0\rangle \\
II &=& \langle 0|\bar D^{3/2}_{-\frac{3}{2}\frac{3}{2}}
(\bar D^3_{30}D^3_{30}) D^{3/2}_{-\frac{3}{2}\frac{3}{2}}|0\rangle \\
III &=&\langle 0|\bar D^{3/2}_{-\frac{3}{2}\frac{3}{2}}
(\bar D^3_{-30}D^3_{-30})D^{3/2}_{-\frac{3}{2}\frac{3}{2}}|0\rangle \\
IV &=& \langle 0|\bar D^{3/2}_{-\frac{3}{2}\frac{3}{2}}(\bar D^3_{00}
D^3_{00})D^{3/2}_{-\frac{3}{2}\frac{3}{2}}|0\rangle \\
\end{array}
\ee
\no To agree with the standard theory we now require
\be
\nabla^\mu\hat\varphi\nabla_\mu\hat\varphi = \partial^\mu\bar\rho
\partial_\mu\bar\rho + \hat g^2\bar\rho^2
[W_-^\mu W_{-\mu} + W^\mu_+ W_{+\mu} + \frac{1}{\cos^2\theta}
Z^\mu Z_\mu]
\ee
\no where
\be
\begin{array}{rcl}
& &\bar\rho = I^{1/2}\rho \\
& &\frac{II}{I} |C_-|^2 = 1 \\
& &\frac{III}{I} |C_+|^2 = 1 \\
& &\frac{IV}{I} |C_3|^2 = 1 \\
\end{array}
\ee

The $I,II,III$ and $IV$ are all explicit functions of $(q,\beta)$. 
 
The
four coefficients $(C_-,C_+,C_3,C_0)$ then follow from (16.6), (16.8) and (15.11):
\be
\begin{array}{rcl}
C_- &=& \left[\left\langle\matrix{6\cr 3\cr}\right\rangle_1 |\beta|^6
\displaystyle\prod^3_1 (1-q_1^{2t}|\beta|^2)\right]^{1/2} \\
C_+ &=& \left[\left\langle\matrix{6\cr 3\cr}\right\rangle_1 |\beta|^6
\displaystyle\prod^2_0 (1-q^{2s}|\beta|^2)\right]^{1/2} \\
C_3 &=& f_3(|\beta|^2)^{-1} \\
C_0 &=& -\frac{1}{f_0(|\beta|^2)} \tan\theta \\
\end{array}
\ee
\no The mass relations between the neutral and charged vectors that
follow from (16.7) are the same as for the standard theory.  In
order that the vertex functions be consistent with these relations,
the vertex factors must be supplied with the $(C_-,C_+,C_3,C_0)$
given by (16.9).

\section{Discussion.}

We have been able to organize a class of data relevant to, but 
also not accessible from the standard theory.  Although this
model does not permit one to calculate absolute masses and
reaction rates, it does provide a simple frame that describes
fermionic spectra and reaction rates, and emerges quite naturally
from the $q$-electroweak theory.  The model may be fine-tuned
and may be useful as a phenomenological model.  To go further at a
deeper level, one must be able to construct an effective field
theory.

The relation of the knot model based on $q$-electroweak to
standard electroweak resembles the relation of the Schr\"odinger
to the Hamilton-Jacobi equation insofar as one adjoins in both
cases a state space not present in the original description.  
Like the wave equation, the knot model may be applied in a
variety of contexts.  When the Schr\"odinger equation is
applied to a single atom, or to molecules or other systems of
arbitrary complexity, it has to be modified by changing the
appropriate parameters.  In hydrogenic systems, for example, the
wave equation is applied with differing values of $Z$ and
$m$.  When the $q$-knot model is similarly applied to answer quite
different questions, such as the masses of the fermions or reaction
rates among them, it also has to be appropriately modified by 
choosing different values of $q$ and $\beta$.  In every case
the same algebra is used just as in every case the same wave
equation is used.

We have computed only relative masses and relative rates.
Given this restriction we find that $q$ and $\beta$ differ
markedly from unity in the expressions for the mass ratios but are
very close to unity in the corresponding expressions for relative
rates.

If one regards $SU_q(2)$ as a fundamental symmetry, it may be
possible to regard $q$ as a new constant with a single value that
comes out differently in different contexts where external
influences such as the gluon field have been ignored.  Alternatively,
$q$ may be regarded as a running coupling constant, where $\beta$ and
$m_n(w,r)$ determine the energy scale.

The model in its present form predicts a fourth generation of
fermions as well as a neutrino mass spectrum.  In applications
to fermionic mass spectra the parameters
of the model ($q$ and $\beta$) have been fixed by two data
$(M,m)$.  If a fourth generation is found or not found in the
neighborhood predicted by the model, then the model can be
refined. 

In the standard theory, and therefore here as well, there is no
attempt to go beyond a provisional expression $(\bar\psi\varphi\psi)$
for the fermionic masses, i.e. $\bar\psi\varphi\psi$ could just as
well be replaced by $\bar\psi$ $F(\varphi)\psi$.  There is 
no difficulty in cutting off the mass spectrum at three generations
without changing the essential structure of the model.

The neutrino mass spectrum is also a strong constraint
on the model; at present the data on this spectrum 
are compatible with $q\cong 1$.
In applications to fermionic currents, both in the lepton-neutrino
sector and in the Kobayashi-Maskawa sector, the data are compatible
with $q\cong 1$.

The form of the vector coupling is restricted in the lepton sector
by lepton conservation and by the Universal Fermi Interaction.  In the
quark sector it is restricted by the Kobayashi-Maskawa matrix.  All of
these restrictions can be satisfied approximately by the simple model described in
this paper, but the model can be refined as more empirical input is
utilized.  Since gluon and gravitational couplings are not explicitly
included, one may tentatively regard the deviation of $q$ from unity as a measure of
their influence.

\vskip.5cm

\no {\bf References.}

\vskip.3cm
\begin{enumerate}
\item R. J. Finkelstein, hep-th/0408218, to appear in Int. J. Mod.
Phys. (A).
\item R. J. Finkelstein, hep-th/0110075, Lett. Math. Phys. {\bf 62}, 199-210 (2002).
\item S. Eidelman {\it et al.}, Phys. Lett. B{\bf 592}, 1 (2004).
\item V. Barger {\it et al.}, Phys. Lett. B{\bf 595}, 55 (2004).
\item A. C. Cadavid and R. J. Finkelstein, J. Math. Phys. {\bf 36},
1912 (1995).
\item Phys. Rev. D Part I, Review of Particle Physics, {\bf 66} (2002). 
\end{enumerate}

\end{document}